\documentclass[aps,prc,twocolumn,preprintnumbers,amsmath,amssymb,10pt,showpacs,superscriptaddress,nofootinbib]{revtex4-1}
\usepackage{xcolor}
\usepackage{epsfig,graphicx,float,pst-all}
\usepackage{mathrsfs}
\usepackage[utf8]{inputenc}
\usepackage{url}
\usepackage[caption = false]{subfig}
\usepackage[normalem]{ulem}
\begin{document}
	\newcommand{\js}[1]{\textcolor{blue}{\it #1}}
	\newcommand{\jsout}[1]{\textcolor{blue}{\sout{#1}}}
	\newcommand{\gs}[1]{\textcolor{red}{\it #1}}
	\newcommand{\gsout}[1]{\textcolor{red}{\sout{#1}}}
	\newcommand{\lf}[1]{\textcolor{magenta}{\it #1}}
	\newcommand{\lfout}[1]{\textcolor{magenta}{\sout{#1}}}
	
	\title{Exploring the halo character and dipole response in the dripline nucleus $^{31}$F} 
	
	
	\author{G. Singh}
	\email{g.singh@unipd.it}
	\affiliation{Dipartimento di Fisica e Astronomia ``G.Galilei'', Università degli Studi di Padova, via Marzolo 8, Padova, I-35131, Italy}
	\affiliation{INFN-Sezione di Padova, via Marzolo 8, Padova, I-35131, Italy}
	\author{Jagjit Singh}
	\email{jsingh@rcnp.osaka-u.ac.jp}
	\affiliation{Research Centre for Nuclear Physics (RCNP), Osaka University, Ibaraki 567-0047, Japan}
	\affiliation{Department of Physics, Akal University, Talwandi Sabo, Punjab, India-151302}
	\author{J. Casal}
	\email{jcasal@us.es}
	\affiliation{Departamento de F\'{i}sica At\'{o}mica, Molecular y Nuclear, Facultad de F\'{i}sica, Universidad de Sevilla, Apartado 1065, E-41080 Sevilla, Spain}
	\author{L. Fortunato}
	\email{fortunat@pd.infn.it}
	\affiliation{Dipartimento di Fisica e Astronomia ``G.Galilei'', Università degli Studi di Padova, via Marzolo 8, Padova, I-35131, Italy}
	\affiliation{INFN-Sezione di Padova, via Marzolo 8, Padova, I-35131, Italy}
	
	\date{\today}

	\begin{abstract}
		\begin{description}
			\item[Background]
			{Lying at the lower edge of the `island of inversion', neutron-rich Fluorine isotopes ($^{29-31}$F) provide a curious case to study the configuration mixing in this part of the nuclear landscape. Recent studies have suggested that a prospective two-neutron halo in the dripline nucleus $^{31}$F could be linked to the occupancy of the $pf$ intruder configurations.}
			\item[Purpose]
			{Focusing on configuration mixing, matter radii and neutron-neutron ($nn$) correlations in the ground-state of $^{31}$F, we explore various scenarios to analyze its possible halo nature as well as the low-lying electric dipole ($E$1) response within a three-body approach.}
			\item[Method]
			{We use an analytical, transformed harmonic oscillator basis under the aegis of a hyperspherical formalism to construct the ground state three-body wave function of $^{31}$F. The $nn$ interaction is defined by the Gogny-Pires-Tourreil potential that includes the central, spin-orbit and tensor  terms, while the different two-body potentials to describe the core + $n$ subsystems are constrained by the different possible scenarios considered.} 
			
			\item[Results]
			The $^{31}$F ground-state configuration mixing and its matter radius are computed for 
			different choices of the $^{30}$F structure coupled to the valence neutron. The admixture of {$p_{3/2}$, $d_{3/2}$, and $f_{7/2}$} components is found to play an important role, favouring the dominance of inverted configurations with dineutron spreads for two-neutron halo formation. The increase in matter radius with respect to the core radius, $\Delta r \gtrsim$ 0.30\,fm and the dipole distributions along with the integrated B$(E1)$ strengths of $\geqslant$ 2.6\,$e^2$fm$^2$ are large enough to be compatible with other two-neutron halo nuclei.
			
			\item[Conclusion]
			{Three-body results for $^{31}$F indicate a large spatial extension in its ground state due to the inversion of the energy levels of the normal shell model scheme. The increase is augmented by and is proportional to the extent of the $p_{3/2}$ component in the wave function. Additionally, the enhanced dipole distributions and large B$(E1)$ strengths all point to the two-neutron halo character of $^{31}$F.}
			
		\end{description}
	\end{abstract}
	
	\maketitle

	
	\section{Introduction}
	\label{sec:intro}
	
	Exotic nuclei have been a field of interest for a long time now, especially because of their 
	peculiar behavior in comparison to the nuclei lying in the valley of stability. New developments in experimental facilities all over the world have enabled researchers to push the limits of knowledge on these exotic systems. Consequently, the interest in these uniquely organized subatomic structures has piqued, gradually encompassing not only the lower mass weakly bound nuclei, but also those in the medium mass region. In particular, the `island of inversion' \cite{WBB90PRC, Brow10Phys} has garnered much attention due to the reduced shell gaps, leading to inversion and possible mixing of the $pf$- and $sd$-shell neutron configurations in F, Ne, Na and Mg isotopes \cite{GMO12PRL, AFG19PRL, BKT20PRL, FCH20CP, SC14NPA, SNC15NPA, MDS21NPA, SSC16PRC, KNK16PRC, NKK09PRL}. The mixing of the $pf$- and $sd$-shell energy levels also leads to possible formation of a halo in some of these isotopes, where the low $l$ configuration ($l=0,1$) favors the extension of the tail of the wave function of the valence neutron(s) to diffuse into the classically forbidden region, giving the nucleus a larger matter radius \cite{JRF04RMP, TSK13PPNP}. In addition, isotopes in this patch of the nuclear chart are important sources of seed nuclei for the initiation of the $r$-process of nucleosynthesis in stellar plasma as well as predicting the abundance of drip nuclei \cite{SKM05AAS, TSK01AAS, SCS17PRC, SSC17PRC}, making it all the more imperative that their properties and behavior be known precisely.
	
	Fluorine and Neon are even more exciting in that recently their dripline nuclei have been well established by the works of Ahn \textit{et al.} \cite{AFG19PRL} and which now have been assigned mass numbers 31 and 34, respectively. In fact, Fluorine isotopes, with $^{29}$F, are also known to form the heaviest Borromean halo confirmed to date \cite{SCH20PRC, BKT20PRL, FCH20CP}. However, speculations are rife about the last bound isotope of the chain, $^{31}$F, which was first discovered in an experiment by Sakurai \textit{et al.} \cite{SLN99PLB}, being a two-neutron halo or even an anti-halo \cite{MLX20PRC, MHK20PRC}. The experiment also confirmed the instability of its isotopic predecessor, $^{30}$F, whereas $^{31}$F was deemed to be bound \cite{MNN95ANDT,SLN99PLB}. The particle stability of $^{31}$F over $^{30}$F could be owed to the neutron-neutron correlation energy of odd-odd fluorine isotopes being higher with respect to the odd-even isotopes. 
	
	A chronological survey suggests that a case of strong configuration mixing was presented for the last two dripline isotopes of Fluorine \cite{PR94NPA}. Further calculations based on the No-Core Shell Model (NCSM) approach convoluted with the Multi-Configurational Perturbation Theory (MCPT) show that for fluorine isotopes the ground state (g.s.) energies remained more or less constant from $^{25}$F to $^{30}$F due to the opening of the $f_{7/2}$ shell \cite{TGV18PLB}.
	For $^{31}$F, subsequent experiments and inspections of the data again confirmed it to be very weakly bound \cite{Pen01PAN, NSA02PLB, LPD03PAN, LP04PAN, PL06PPN}. Analysis of the $^9$Be($^{32}$Mg,$^{30}$Ne) reaction using the SPDF-M interaction in Monte-Carlo Shell Model \cite{UOM01PRC} revealed an \textit{ad hoc} lowering of the $1f_{7/2}$ and $2p_{3/2}$ single particle energies by 800\,keV was required in the simulations to establish the bound nature of $^{31}$F. 
	Ensuing investigations have indeed conveyed that the ground and first excited states of this dripline nucleus should be largely mixed \cite{CNP14PRC}, while its density distribution is quite extended with respect to $^{29}$F which makes it a possible halo at the outer surface \cite{CPF20PRC}. This is intriguing because its supposed core ($^{29}$F) is itself a halo, and so far, no halo with a halo core has been discovered yet in the island of inversion.

	The concept of the `pairing anti-halo effect' has also been introduced in $^{31}$F,
	which is the reduction in the spatial expansion of the nucleus in comparison to the usual mean field estimates \cite{MHK20PRC, HS11PRC, BDP00PLB}. This decrease in the matter radius results from the modification of the asymptotics of the valence neutron wave functions by the pair potential of the neutrons, which may be present, for example, in a Borromean nucleus. Separately, a mainly qualitative description pushing for strongly prolate fluorine isotopes due to the mixing of the $pf$-shells suggests that the halo character of $^{31}$F should depend resolutely on the size of the deformation present in the system \cite{HAMA21PLB}. It is deemed to be quite independent of the amount of neutron correlation, which is surprising because calculations using the Gamow Shell Model (GSM) with effective field theory (EFT) and Furutani-Horiuchi-Tamagaki (FHT) interactions have shown that it is the two valence neutrons (much more delocalized than the core of the nucleus) that give the main contribution towards the extended root-mean-square (rms) radius and correlation density 
	since the proton part of the wave functions is highly localized. It is argued that the considered two neutron separation energy, $S_{2n}$, of 0.17\,MeV is sufficiently small to sustain a halo in $^{31}$F \cite{MLX20PRC}, provided one considers the g.s. to be composed of the $p_{3/2}$ partial wave for the two valence neutrons. Although the $2n$-halo seems to be made up of a $p_{3/2}$ orbit, the contributions from the $d_{3/2}$, and $f_{7/2}$ states is not ruled out entirely.

	Evidently, the g.s. structure of $^{31}$F and the quantification of its halo nature are open questions that need to be addressed. Therefore, in this work, we explore the different situations and configurations that could modify the structure of this dripline nucleus and compute various observables like the matter radius, density distributions including also the dipole response, sum rules, etc., through the generation of the requisite energy spectra in the different scenarios considered. We also try and shed light on whether $^{31}$F presents the fascinating and uncharted possibility of a halo nucleus with a halo core in this region of the nuclear chart (although we consider an inert core in the present study) or the feasibility of an anti-halo effect. We analyze the possible ground state configuration and halo character of $^{31}$F, employing a three-body formalism in hyperspherical coordinates \cite{NFJ01PR}. In the pseudostates (PS) method \cite{TON97PRL} that we implement using the transformed harmonic oscillator (THO) basis, the bound states are manifested through the eigenstates corresponding to the negative energy eigenvalues, while the discretized continuum is illustrated by the positive energy states. The approach has been applied successfully in the past to other Borromean nuclei to study their structural as well as reaction dynamics \cite{CRA14PRC, SCH20PRC, Casal18PRC}. The limited information available from previous theoretical and experimental works is utilized to refine the various potential parameter sets and to scrutinize the subtlety of $^{31}$F behavior to the low lying continuum of $^{30}$F.
	
	\color{black}
	The next section presents our formalism, while we discuss our results in Section \ref{sec: R&D} and conclude in Section \ref{sec: Conc}.
	
	
	
	\section{Formalism}
	\label{sec:For}
	
\begin{figure}[htbp]
		\centering
		\includegraphics[trim={1cm 1cm 1cm 1cm},clip,width=8.3cm]{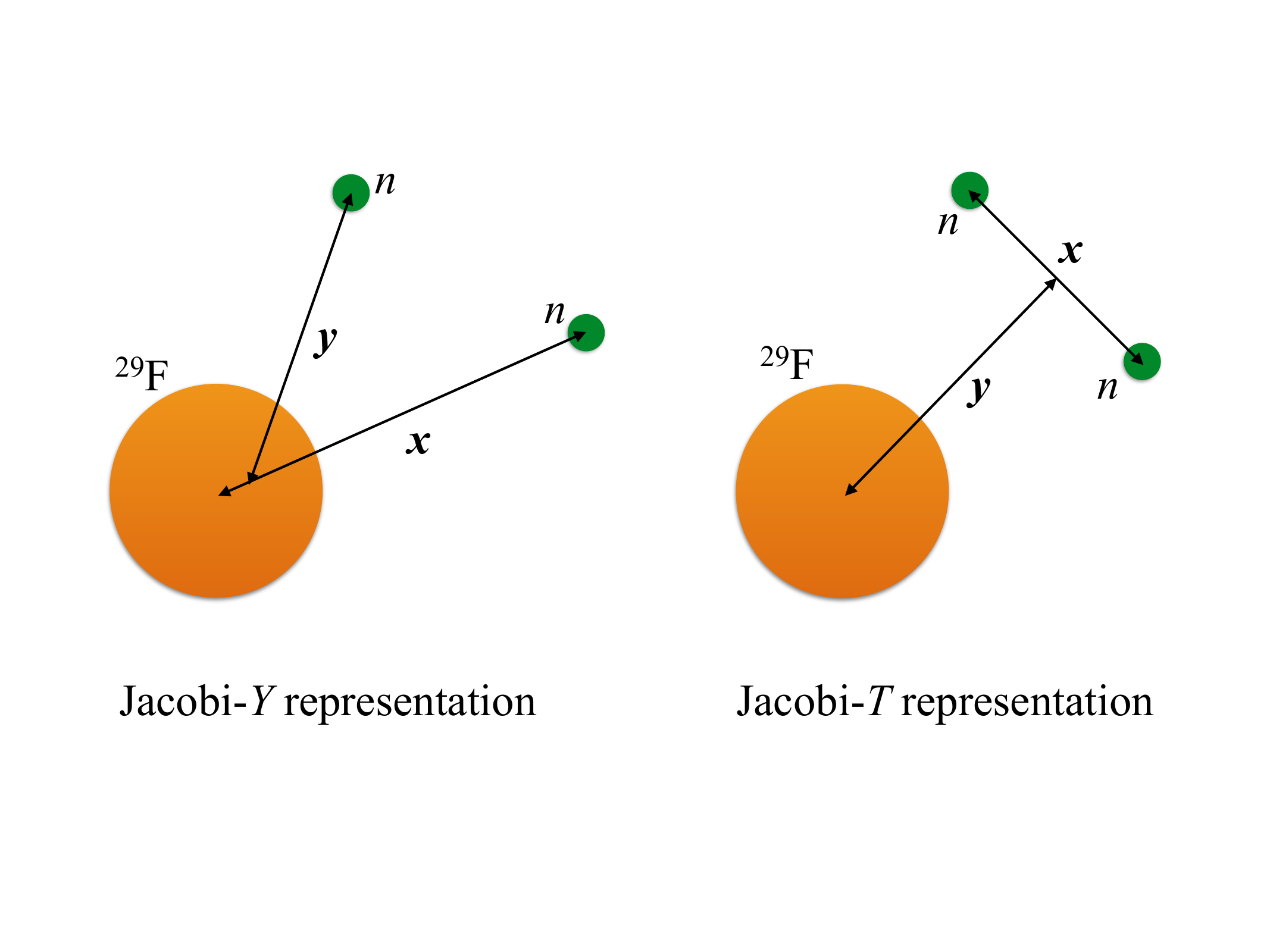}
		
		\caption{\label{fig: JacobiY&T}  The Jacobi-$Y$ and Jacobi-$T$ coordinate representations for a $^{29}$F + $n$ + $n$ system comprising the three-body composite $^{31}$F.}
	\end{figure}

	The Hamiltonian eigenstates of a three-body system, in accordance with the hyperspherical formalism can be expanded as \cite{ZDF93PR, SCH20PRC, CRA14PRC, Casal18PRC},
	\begin{equation}
	\psi^{jm_j}(\rho,\Omega) = \frac{1}{\rho^{5/2}}\sum_{\beta}R_{\beta}^{j}(\rho)\mathcal{Y}_{\beta}^{jm_j}(\Omega),
	\label{eq:3bwf}
	\end{equation}
	where $\rho^2 (= x^2 + y^2)$ is the hyperradius specified using the Jacobi coordinates $\{\boldsymbol{x},\boldsymbol{y}\}$, while the angular dependence is established through $\Omega\equiv\{\alpha,\widehat{x},\widehat{y}\}$; $\alpha=\tan^{-1}(x/y)$ being the hyperangle \cite{CSF20PRC}. A transformation between a Jacobi-$T$ and a Jacobi-$Y$ coordinate scheme (see Fig. \ref{fig: JacobiY&T}) changes this hyperangle, while the hyperradius $\rho$ is preserved. Although both the representations of the Jacobi coordinates are equivalent, the Jacobi-$T$ set is better suited for a core + $n$ + $n$ system as the wave function must be antisymmetric under the exchange of identical particles connecting the \textit{\textbf{x}} coordinate.
	
	For each channel $\beta$, $R_{\beta}(\rho)$ and $\mathcal{Y}_{\beta}(\Omega)$ in Eq. (\ref{eq:3bwf}) are the hyperradial functions and the hyperangular functions respectively, with $\beta$ denoting a set of quantum numbers coupled to the total angular momentum $j$ such that, $\beta \equiv \left\lbrace K, l_x, l_y, l, S_x, J_{ab}\right\rbrace j$. For two particles coupled by the \textit{\textbf{x}} coordinate having a total spin $S_x$, $\boldsymbol{J}_{ab}=\boldsymbol{l}+\boldsymbol{S}_x$, where the orbital angular momenta follow, $\boldsymbol{l}=\boldsymbol{l}_x+\boldsymbol{l}_y$. $K$ is the hypermomentum and also represents an effective three-body barrier in the description of the kinetic energy part of the set of coupled Schr\"{o}dinger equations \cite{SCH20PRC, Casal18PRC},
	\begin{equation}
	\begin{split}
	&\left[-\frac{\hbar^2}{2m}\left(\frac{d^2}{d\rho^2}-\frac{15/4+K(K+4)}{\rho^2}\right)-\textrm{E}\right]R_{\beta}^{j}(\rho)\\&+\sum_{\beta'}V_{\beta'\beta}^{jm_j}(\rho) R_{\beta'}^{j}(\rho)=0,
	\end{split}
	\label{eq:coupled}
	\end{equation}
	with $m$ denoting a scaling mass, typically taken to be that of a nucleon. Above, $V_{\beta'\beta}^{jm_j}(\rho)$ are the coupling potentials defined as,
	\begin{equation}
	V_{\beta'\beta}^{jm_j}(\rho)=\left\langle \mathcal{Y}_{\beta }^{jm_j}(\Omega)\Big|V_{12}+V_{13}+V_{23} \Big|\mathcal{Y}_{\beta'}^{ jm_j}(\Omega) \right\rangle,
	\label{eq:3bcoup}
	\end{equation}
	where $V_{mn}$, with ($m,n \in \left\lbrace 1, 2, 3 \right\rbrace $), are the two-body potentials between the pairs of particles within the three-body composite and are described in detail later. The $\mathcal{Y}_{\beta}^{ jm_j}(\Omega)$ in the above equation as well as in Eq. (\ref{eq:3bwf}) are states of good angular momentum $j$, that follow the coupling scheme with
	\begin{equation}
	\mathcal{Y}_{\beta}^{jm_j}(\Omega)=\left\{\left[\Upsilon_{Kl}^{l_xl_y}(\Omega)\otimes\varphi_{S_x}\right]_{J_{ab}}\otimes\chi_I\right\}_{jm_j}.
	\label{eq:Upsilon}
	\end{equation}
	Here, the $\Upsilon_{Kl}^{l_xl_y}(\Omega)$ are the hyperspherical harmonics (HH) that are the eigenstates of the hypermomentum operator $K$ \cite{ZDF93PR} and are described for a given $(l_x,l_y)_l$ coupling through the usual spherical harmonics $Y^{m_l}_l$ via,
	\begin{equation}
	\Upsilon_{Klm_l}^{l_xl_y}(\Omega)=\phi_K^{l_xl_y}(\alpha)\left[Y_{l_x}(\boldsymbol{x})\otimes Y_{l_y}(\boldsymbol{y})\right]_{lm_l};
	\label{eq:HH}
	\end{equation}
	\begin{equation}
	\begin{split}
	\phi_K^{l_xl_y}(\alpha) = N_{K}^{l_xl_y}\left(\sin\alpha\right)^{l_x}\left(\cos\alpha\right)^{l_y}P_n^{l_x+\frac{1}{2},l_y+\frac{1}{2}}\left(\cos 2\alpha\right),
	\label{eq:varphi}
	\end{split}
	\end{equation}
	with $N_K^{l_xl_y}$ being a normalization constant and $P_n^{\mu,\nu}$ being a Jacobi polynomial of order $n = (K - l_x - l_y)/2$. Notice that if the core spin $I$ is disregarded ($I=0$) to simplify the designated core +  $n$ potential, we are then able to contract the expression in Eq. (\ref{eq:Upsilon}) as,
	\begin{equation}
	\mathcal{Y}_{\beta}^{jm_j}(\Omega)=\left[\Upsilon_{Kl}^{l_xl_y}(\Omega)\otimes\chi_{S_x}\right]_{jm_j},
	\label{eq:Upsilon0}
	\end{equation}
	where we have used $\boldsymbol{J}_{ab}=\boldsymbol{j}$. 
	This intentional neglect of the core spin for an odd-even nucleus is not new and is known to have been employed with considerable success, for instance, in $^{11}$Li and $^{29}$F \cite{SCH20PRC,MTO19PTEP}. However, care must be taken for other quantum numbers. For example, the Pauli principle dictates that for two identical neutrons, as is the case with $^{31}$F, a condition for isospin $T$ = 1 necessarily means that $S_x + l_x$ should always be an even number. Further, since $S_x$ is the spin of the two-neutron pair, each having spin 1/2, therefore, any value of $S_x$ different from 0 or 1 is forbidden.
	
	Meanwhile, the solutions to the set of coupled equations displayed by Eq. (\ref{eq:coupled}) are represented by the hyperradial functions that can be expanded in a discrete basis through,
	\begin{equation}
	R_\beta^{j}(\rho)=\sum_{i}^{i_{max}} C_{i\beta}^j U_{i\beta}(\rho)
	\label{eq:radial}.
	\end{equation}
	At this point, the expansion coefficients $C_{i\beta}^j$ can easily be obtained by diagonalizing the three-body Hamiltonian for $i = 0, 1, ... i_{max}$ basis functions or the hyperradial excitations. The basis preference is not unique and several options are available within the PS method \cite{RAG05PRC,DDB03PRC,MHY04NPA}. We choose the transformed harmonic oscillator basis \cite{KAG05PRC,Casal18PRC,CRA13PRC}, generated through the harmonic oscillator (HO) functions in the hyperspherical coordinates via,
	\begin{equation}
	U_{i\beta}^{\text{THO}}(\rho)=\sqrt{\frac{du}{d\rho}}U_{iK}^{\text{HO}}[u(\rho)],
	\label{eq:THO}
	\end{equation}
	where,
	\begin{equation}
	U_{iK}^{\text{HO}}(u)=N_{iK}u^{K+5/2}L_i^{K+2}(u)\exp{\left(-u^2/2\right)},
	\label{eq:fHO}
	\end{equation} 
	with $N_{iK}$ being the normalization constants and $L_i^{K+2}(u)$ being the generalized Laguerre polynomials. The local scale transformation through the $u(\rho)$ functions in Eq. (\ref{eq:THO}) is defined by,
	\begin{equation}
	u(\rho) = \frac{1}{b\sqrt{2}}\left[\frac{1}{\left(\frac{1}{\rho}\right)^{4} +
		\left(\frac{1}{\gamma\sqrt{\rho}}\right)^4}\right]^{\frac{1}{4}},
	\label{eq:LST}
	\end{equation}
	depending upon two parameters: $b$ and $\gamma$. Their ratio, $\gamma/b$, regulates the PS density after diagonalization by governing the hyperradial extension of the basis \cite{Casal18PRC, CRA13PRC, CSF20PRC}. A smaller $\gamma$ for a given $b$ is required to administer a larger number of pseudostates just above the threshold of the continuum, a suitable and preferred arrangement when computing, for instance, electromagnetic transitions. The two parameters are also vital to convert the Gaussian asymptotic behavior of the harmonic oscillator functions into a simple exponential form, i.e., $e^{-\gamma^2\rho/2b^2}$, a transformation that improves the convergence of the calculations with respect to the number of basis functions $i_{max}$.

	\subsection{Two-body potentials}
	Since we deal also with the continuum, the spectral properties of core$~+~n$ subsystems play a crucial role in the structure of Borromean nuclei, as the corresponding potential enters explicitly in the three-body Hamiltonian (see Eq (\ref{eq:3bcoup})). To that effect, we describe here the parameters used for calculations done to understand $^{31}$F, considering it as a three-body system of $^{29}$F and two valence, correlated neutrons. Taking $^{29}$F to be an inert and spinless core, there are three, two-body interactions in the system, viz., neutron-neutron or $nn$, $^{29}$F-$n$ and $n$-$^{29}$F. In this work, the $nn$ interaction is described by the Gogny-Pires-Tourreil (GPT) potential \cite{GPT}. This is a tensor potential that reproduces rather well the low-energy $nn$ scattering data. Other nucleon-nucleon interactions are also available in the literature (e.g., the tensor force in Ref. \cite{GFJ04PRC}, or the Minnesota potential \cite{TLT77NPA}), but the choice of the interaction is not crucial in terms of wave-function properties, provided a realistic $nn$ potential is employed~\cite{SCH20PRC}.  
	The $\text{core}+n$ interaction is modeled with central and spin-orbit terms~\cite{BMBook},
	\begin{equation}
	V_{({\rm ^{29}F}+n)} = -V_0 f(r) + V_{ls}\lambda_\pi^2\frac{1}{r}\frac{df(r)}{dr} \vec{l}\cdot\vec{s},
	\end{equation}
	where $f(r)$ has the standard Woods-Saxon form
	\begin{equation}
	f(r)=\frac{1}{1+{\rm\,exp}\left(\frac{r-R_c}{a}\right)}.
	\end{equation}
	Here, $R_c = r_0A_c^{1/3}$, with $r_0$ being the radius parameter fixed at $1.25$\,fm and $A_c$ being the core mass, which for the present case is $29$. The diffuseness $a$, is set to 0.75\,fm and $\lambda_\pi$ = 1.414\,fm, is the Compton wavelength. These parameters have been used successfully in the past to study the structural aspects of $^{29}$F \cite{SCH20PRC, CSF20PRC}. The spin-orbit potential depth follows the systematics,
	$V_{ls}=\left(22-14\frac{(N-Z)}{A}\right)$ \cite{BMBook}, which for the current case, evaluates to, $V_{ls}$ = 16.690\,MeV.
	
	The depth of the central potential, $V_0$, which in general, is $l$-dependent, is adjusted to fix the various configurations considered in this study 
	for the $^{29}$F$~+n$ two-body subsystem. The $l$-dependence of this central potential depth allows to incorporate effects beyond the inert core$~+~n$ picture. The values of $V_0$ are presented in Table \ref{T_pot} for each of the case with different sets considered to contribute to the ground state of $^{31}$F. $E_R$ is the resonance energy obtained in the continuum of an unbound $^{30}$F, that was made to mimic the case A of Table I of Ref. \cite{MHK20PRC} for the largest radius configuration. 
	
	\begin{table}[htbp]
		\caption{\label{T_pot} Description of interactions considered for the $^{29}$F$~+~n$ systems. $V_0$ is the central Woods-Saxon potential depth and $E_R$ is the resonance energy in the continuum of unbound $^{30}$F. The parameters, $r_0$ = 1.25\,fm, $a$ = 0.75\,fm and spin-orbit potential depth, $V_{ls}$ = 16.690\,MeV, are kept constant. Cases I and II refer to the closed shell filling of $^{29}$F with four valence neutrons in $1d_{3/2}$ and $2p_{3/2}$ subshells, respectively. Case III accounts for the open shell structure. The sets A, B and D account for normal shell-model, intruder and inverted configurations, respectively. Set C is similar to set B, but with the inversion of the $1d_{3/2}$ and $2p_{3/2}$ subshells, while sets AH and AH2 in cases I and III represent the possible scenarios of the anti-halo effect. For details, please see text.}
		\centering
		\begin{tabular}{ccccc} 
			\toprule\\[-1.5ex]
			Case & Set & $lj$ & $V_0$ (MeV) & $E_R$ (MeV)   \\ [1ex]
			\colrule\hline\\[-1.5ex]
			\multicolumn{5}{c}{\textbf{Closed $\boldsymbol{(1d_{3/2})^4}$ shell}}\\ [1ex]
			\colrule\hline\\[-1.5ex]
			I & A & $f_{7/2}$ & 46.490 & 0.23 \\
			&  & $p_{3/2}$ & 42.726 & 1.21 \\
			& & & & \\
			I & D & $p_{3/2}$ & 45.230 & 0.14 \\
			&  & $f_{7/2}$ & 43.920 & 1.22 \\
			& & & & \\
			I & AH & $p_{3/2}$ & 45.230 & 0.14 \\
			& & $f_{7/2}$ & 46.490 & 0.23 \\ [1ex]
			\colrule\hline\\[-1.5ex]
			\multicolumn{5}{c}{\textbf{Closed $\boldsymbol{(2p_{3/2})^4}$ shell}}\\ [1ex]
			\colrule\hline\\[-1.5ex]
			II & A & $d_{3/2}$ & 37.374 & 0.14 \\
			&  & $f_{7/2}$ & 43.920 & 1.22 \\
			& & & & \\
			II & D & $f_{7/2}$ & 46.490 & 0.23 \\
			&  & $d_{3/2}$ & 34.785 & 1.22\\ [1ex]
			\colrule\hline\\[-1.5ex]
			\multicolumn{5}{c}{\textbf{Open $\boldsymbol{(1d_{3/2})^2(2p_{3/2})^2}$ shells}} \\ [1ex]
			\colrule\hline\\[-1.5ex]
			III & A & $d_{3/2}$ & 37.805 & 0.11 \\
			&  & $f_{7/2}$ & 46.490 & 0.23 \\
			&  & $p_{3/2}$ & 42.726 & 1.21 \\
			& & & & \\
			III & B & $d_{3/2}$ & 37.805 & 0.11 \\
			&  & $p_{3/2}$ & 45.230 & 0.14 \\
			&  & $f_{7/2}$ & 43.920 & 1.22 \\
			& & & & \\
			III & C & $p_{3/2}$ & 45.414 & 0.11 \\
			&  & $d_{3/2}$ & 37.374 & 0.14 \\
			&  & $f_{7/2}$ & 43.920 & 1.22 \\
			& & & & \\
			III & D & $p_{3/2}$ & 45.414 & 0.11 \\
			&  & $f_{7/2}$ & 43.920 & 1.22 \\
			&  & $d_{3/2}$ & 32.750 & 2.00 \\
			& & & & \\
			III & AH & $p_{3/2}$ & 45.230 & 0.14 \\
			&  & $f_{7/2}$ & 46.490 & 0.23 \\
			&  & $d_{3/2}$ & 34.785 & 1.22 \\
			& & & & \\
			III & AH2 & $d_{3/2}$ & 37.805 & 0.11 \\
			&  & $p_{3/2}$ & 45.230 & 0.14 \\
			&  & $f_{7/2}$ & 46.490 & 0.23 \\ [1ex]
			\botrule
		\end{tabular}
	\end{table}
	
	
	
	Due to a paucity of experimental information on the low lying spectrum of $^{30}$F, we scrutinize three possible cases for our study: Cases I and II refer to the scenarios where a \textit{closed} shell configuration of $^{29}$F would couple with the valence neutron to form the low lying continuum of the $^{30}$F system, i.e., $^{29}$F is a core with fully filled shells. In case I, we have the closed shell configuration for inert $^{29}$F where we put its four valence neutrons in the $1d_{3/2}$ state, making it unavailable for the extra neutron of $^{30}$F. In case II, we place these neutrons in the $2p_{3/2}$ state, pushing it lower in energy than the $1d_{3/2}$ level and opening the latter to accommodate the valence neutron of $^{30}$F. Sets A and D then present the normal shell model ordering and the inverted schemes to fill the valence neutron in $^{30}$F. As an example, case II, set D would cater to the situation where the $2p_{3/2}$ shell is fully filled and the low-lying continuum of $^{30}$F is formed by the $f_{7/2}$ state at a resonance energy of 0.23\,MeV above the threshold,
	thus coming below the $d_{3/2}$ state which is fixed at 1.22\,MeV. Hence, the name `inverted' configuration as compared to the normal shell model ordering. {It must be noted that to avoid numerical instabilities at low energies due to very narrow states associated with high centrifugal barriers, we adopted a slightly larger $f_{7/2}$ energy of 0.23\,MeV as compared to the values of lowest state energies for the $p_{3/2}$ or $d_{3/2}$ levels used in other configurations.}
	
	In case III, we consider the \textit{open} shell core configuration where the four valence neutrons of $^{29}$F go in $(1d_{3/2})^2(2p_{3/2})^2$ orbitals. Sets A and D then describe the situations as averred above. Set B covers the condition when the $p_{3/2}$ level is much lower in energy than the $f_{7/2}$ level, but still slightly above the $d_{3/2}$ energy state. We call this state of affair as the `intruder' configuration as one of the $pf$-levels disentangles from the usual $pf$-shell ordering and comes very close in energy to the $sd$-shell by reducing the shell gap (associated with $N~=~20$), but the `inversion' between the $pf$- and $sd$-shells does not occur. This is a distinction similar to the one made in Ref. \cite{SCH20PRC} and, in conjunction with set C, enables us to explore the interplay between the typical $pf$- and $sd$-shells. Set C is similar to set B, except that now the $d_{3/2}$ and $p_{3/2}$ shells invert their positions, with the $f_{7/2}$ shell still lying much higher in energy. We call it the `\textit{partially} inverted' configuration. Finally, sets AH in cases I and III explore the recently proposed scheme of the anti-halo effect in $^{31}$F. Here, the $f$- and $p$-levels are closer in energy leading to the possibility of contribution by a more stable pairing energy configuration. Note that for set AH in case III, we have fixed the $p_{3/2}$ and the $f_{7/2}$ resonances at exactly the same energies as in case I, but with the addition of the higher lying $d_{3/2}$ resonance due to the opening of the shells. Similarly, set AH2 mimics set B,  but explores the prospects of an anti-halo configuration with the $d_{3/2}$ state occupying the lowest energy in the resonant continuum. Meanwhile, the $s$-wave potential was fixed at $34.785$\,MeV for sets D and AH of case III, and at $37.734$\,MeV for all the other scenarios.
	
	\begin{figure}[htbp]
		\centering
		\includegraphics[trim={0 0 0 0},clip,width=8.3cm]{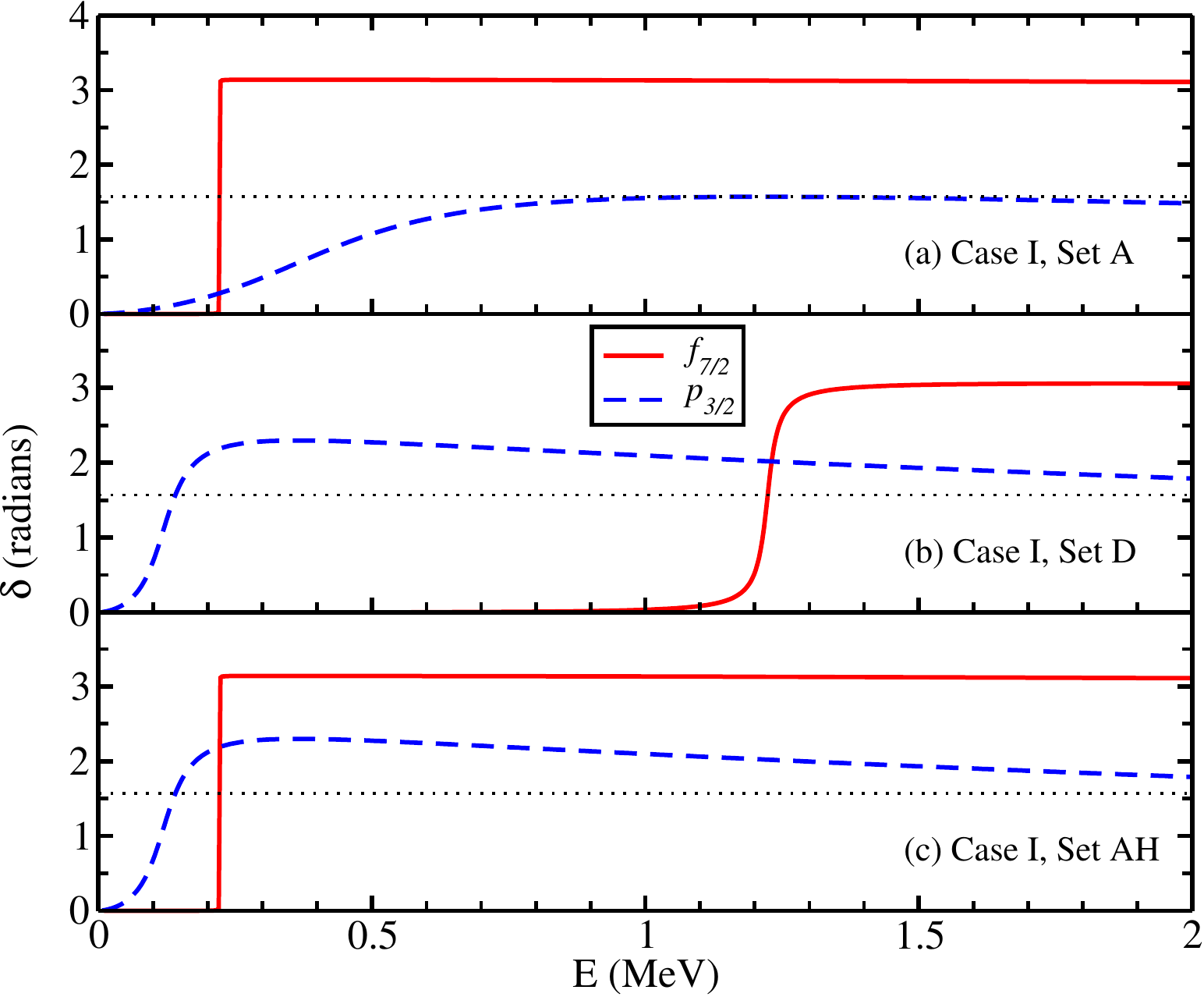}
		\caption{\label{fig: PS_I}  Phase shifts for $^{30}$F corresponding to sets A, D, and AH of case I in panels (a), (b), and (c), respectively. The phase shift curves for the $f_{7/2}$ state are shown by the (red) solid line while the $p_{3/2}$ orbital is represented by the (blue) dashed line. The dotted (black) line corresponds to $\pi/2$ radians.} 
	\end{figure}
	
	\begin{figure}[htbp]
		\centering
		\includegraphics[trim={0 0 0 0},clip,width=8.3cm]{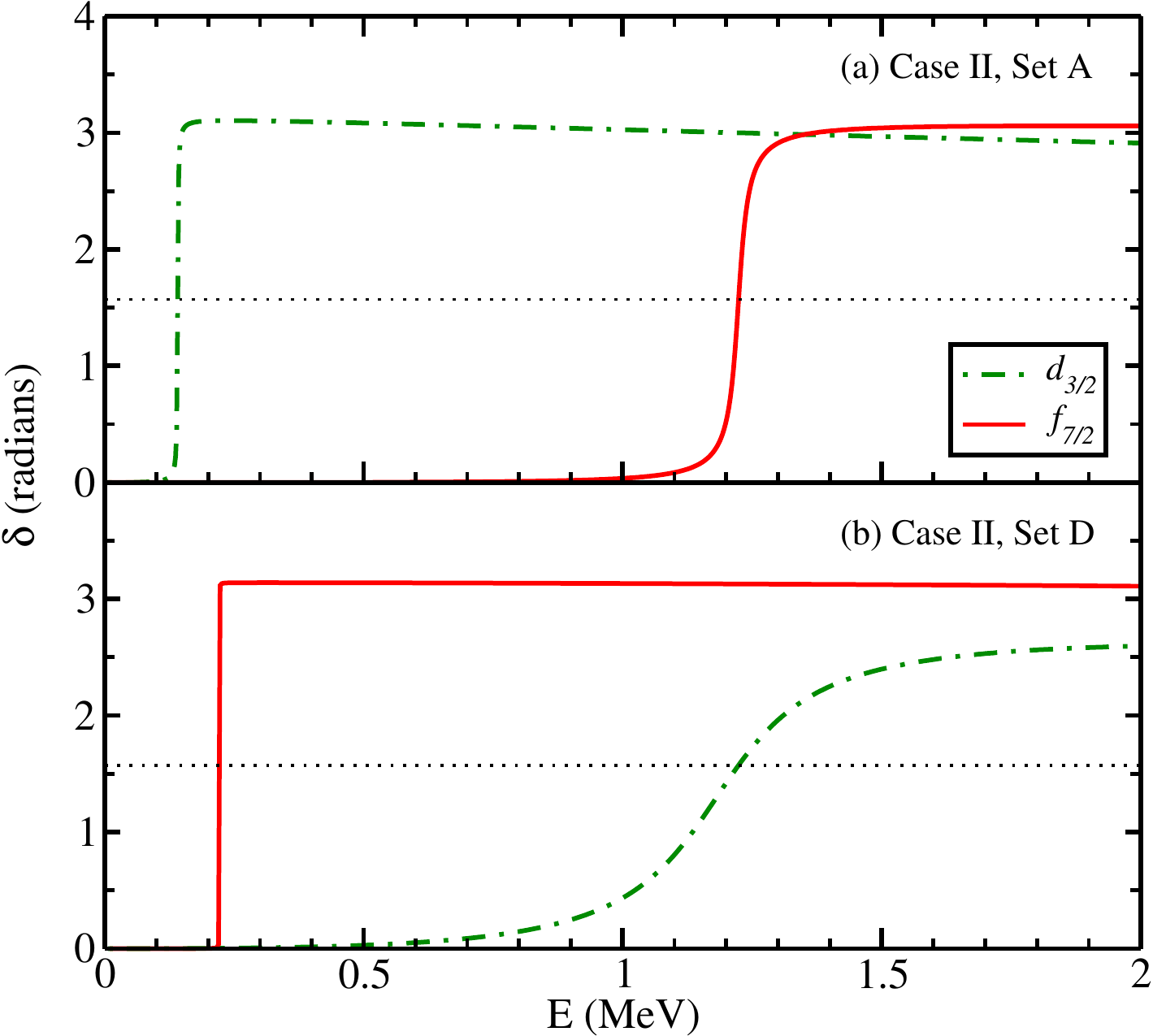}
		\caption{\label{fig: PS_II} Phase shifts for $^{30}$F corresponding to sets A and D of case II in panels (a) and (b), respectively. The phase shift curves for the $f_{7/2}$ orbital are shown by the (red) solid line while the $d_{3/2}$ orbital is represented by the (green) dash-dotted line. The dotted (black) line corresponds to $\pi/2$ radians.} 
	\end{figure}
\begin{figure}[htbp]
		\centering
		\includegraphics[trim={0 0 0 0},clip,width=8.3cm]{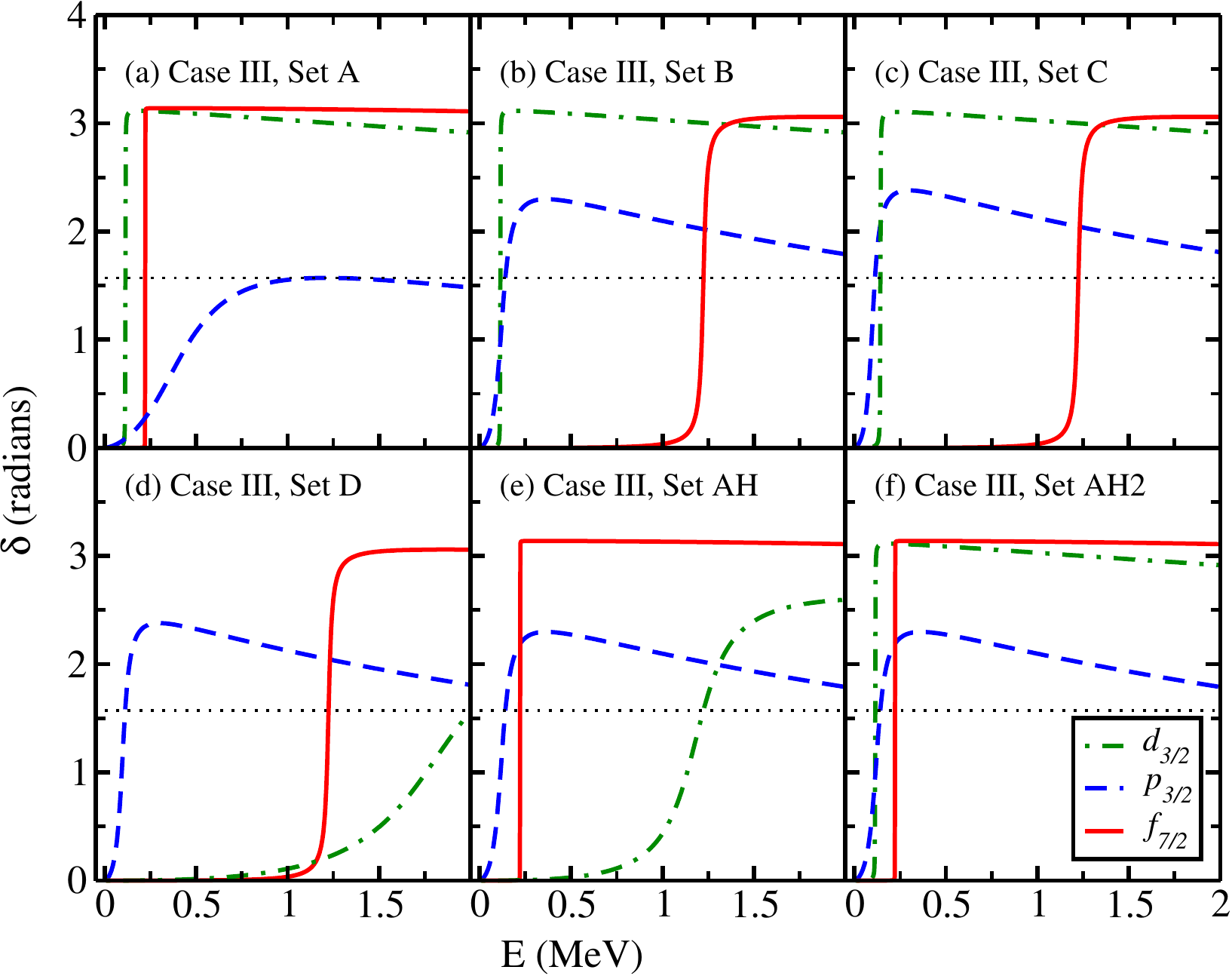}
		\caption{\label{fig: PS_III}  Phase shifts for $^{30}$F corresponding to sets A, B, C, D, AH, and AH2 of case III in panels (a), (b), (c), (d), (e) and (f), respectively. The phase shift curves for the $f_{7/2}$ orbital are shown by the (red) solid line while the $d_{3/2}$ and $p_{3/2}$ orbitals are represented by the (green) dash-dotted and the (blue) dashed lines, respectively. The dotted (black) line corresponds to $\pi/2$ radians.}
	\end{figure}

	In Figs. \ref{fig: PS_I} and \ref{fig: PS_II}, we show the phase shifts for the low lying resonances of $^{30}$F corresponding to sets A, D, and AH for case I and sets A and D for case II, respectively. Similarly, Fig. \ref{fig: PS_III} displays the phase shifts for the six sets pertaining to case III. {For the uninitiated, resonance energy, $E_R$, would be the energy where the phase shifts cross $\pi$/2 radians.}
	
	It may be seen that under the scenarios considered, the $f$-waves present rather sharp resonances, and the numerical mesh size had to be decreased to catch them appropriately. The $d$-wave resonances in the inverted configurations (sets D of cases II and III) are broad as the continuum was so constructed that the $f$- and the $p$-wave resonances lie closer to the threshold.
	It is worth mentioning here that during the search for these two-body potentials to reproduce the positive energy ground state of $^{30}$F in all the configurations for all the cases, it was ensured that all the Pauli forbidden states (that would encourage the presence of unphysical eigenstates of the three-body Hamiltonian) are removed from the valence neutron space. This was achieved by building phase-equivalent shallow potentials within a supersymmetric transformation \cite{Baye87PRL}, which in other words is the spectral mimicry of the equivalent potentials without the bound states. This method had been applied with considerable success in the past \cite{CSF20PRC,SD20PLB} and although other approaches like the Feshbach projection method \cite{FESH62AP,CK13PRL,TTT04NPA,NCT96NPA} exist, a comparison between different treatments of the Pauli states goes beyond the scope of the present work. The interested reader is referred to Ref. \cite{TDE00PRC} for a comparative study of Pauli forbidden states by various approaches.
	
	With this picture, we now present the results in the next section.

	\section{RESULTS AND DISCUSSION}
	\label{sec: R&D}

	We adopt the two neutron separation energy, $S_{2n}$, of $^{31}$F to be 0.150\,MeV \cite{Wang17CPC}, unless specified otherwise. The spin parity of $^{31}$F was taken to be $0^+$ in the ground state as we consider only the neutron degrees of freedom and neglect the proton spin contribution to simplify the construction of the three-body wave function\footnote{Of course, to get the full spin, one should couple the neutron part with that of the unpaired proton.} \cite{SCH20PRC,MTO19PTEP}. The $^{29}$F core matter radius was taken at the central experimental value of 3.50\,fm \cite{BKT20PRL}.
	
	\subsection{Ground state structure}
	\label{sec: g.s.}
	In principle, fixing the two-body potentials does not exactly reproduce the ground state energy of a three-body system, and we require a phenomenological three-body interaction to account for any possible deviations due to effects not incorporated explicitly in the three-body formalism \cite{SCH20PRC,CRA14PRC,TND04CPC}. We take this ancillary potential in a simple Gaussian form:
	
	\begin{equation}
	V^{\rm 3b}(\rho) = V^{3b}_0 \exp[-(\rho/r^{3b}_0)^2].
	\label{3b}
	\end{equation}
	
	\begin{table}[htbp]
		\caption{\label{T_obs} Values of the various observables obtained during the reproduction of the g.s. two-neutron separation energy ($S_{2n}$ = 0.150\,MeV) of $^{31}$F in different schemes of our analysis. The radius parameter $r^{3b}_0$ of the three-body force was kept fixed at 6.0\,fm while the depth of the three-body potential $V^{3b}_0$ ranged from 5 - 10\,MeV depending on the configuration. The matter radius $r_m$, $nn$ distance $r_{nn}$, and core-$nn$ distance $r_{c-nn}$, are listed. $\Delta r$ depicts the change in matter radius with respect to the matter radius of the $^{29}$F core, which was taken to be 3.50\,fm \cite{BKT20PRL}. The sum rules for the $E1$ transitions to the continuum from the $0^+$ g.s. are further mentioned.}
		\centering
		\begin{tabular}{ccccccccc}
			\toprule\\[-1.5ex]
			Case & Set & $r_m$ & $r_{nn}$ & $r_{c-nn}$ & $\Delta r$ & $E1$ Sum rule \\
			& & (fm) & (fm) & (fm) & (fm) & ($e^2$fm$^2$) \\ [1ex]
			\colrule\\[-1.5ex]
			I & A & 3.695 & 6.881 & 4.881 &  0.195 & 1.929 \\
			& D & 3.921 & 9.640 & 6.340 &  0.421 & 3.254 \\
			& AH & 3.861 & 8.963 & 5.986 &  0.361 & 2.902 \\ [1ex]
			\colrule\botrule\\[-1.5ex]
			II & A & 3.617 & 5.728 & 4.270 &  0.117 & 1.477 \\
			& D & 3.599 & 5.732 & 4.017 &  0.099 &1.306  \\ [1ex]
			\colrule\botrule\\[-1.5ex]
			III & A & 3.699 & 6.264 & 4.901 &  0.199 & 1.945 \\
			& B & 3.839 & 8.341 & 5.995 &  0.339 & 2.910 \\
			& C & 3.865 & 8.698 & 6.125 &  0.365 & 3.038 \\
			& D & 3.907 & 9.488 & 6.253 &  0.407 & 3.166 \\
			& AH & 3.823 & 8.400 & 5.793 &  0.323 & 2.718 \\
			& AH2 & 3.801 & 7.905 & 5.742 &  0.301 & 2.670  \\ [1ex]
			\colrule\botrule\\[-1.5ex]
		\end{tabular}
	\end{table}
	
	{For our purpose, we varied the three-body interaction depth, $V^{3b}_0$, to reproduce the two neutron separation energy of 0.150\,MeV for different configurations in the ground state of $^{31}$F. The value of the interaction potential ranged from 5 - 10\,MeV depending on the configuration considered, whereas the radius parameter $r^{3b}_0$, for the interaction, was kept constant at 6.0\,fm \cite{SCH20PRC}.}
	
	{Table \ref{T_obs} shows the variation of the matter radius $r_m$, the distance between the two valence neutrons $r_{nn}$, the distance between the core and the center of mass of the two valence neutrons $r_{c-nn}$, as well as some of the observables computed during this analysis. 
		The physical distances $r_{nn}$ and $r_{c-nn}$ are related to the Jacobi coordinates via, $r_{nn} = x\sqrt{2}$ and $r_{c-nn} = y\sqrt{((A_c+2)/2A_c)}$, $A_c$ being the mass of the core. The sum rules for $E1$ transitions (please see Section \ref{sec: dipole} and the Appendix) from the $0^+$ g.s. to the $1^-$ states in the continuum are also reported here. We compute the change in matter radius of the three-body $^{31}$F with respect to the matter radius of the $^{29}$F core and denote it by $\Delta r$. It is seen to be largest for sets D of cases I and III, indicating strong halo formation possibility. More details about these observables and their properties can be found in Ref. \cite{SCH20PRC}.}
	
	To notify the reader, we mention here that the choice of the three-body potential used in this work is not unique and one can also tune the binary potentials using scaling factors to obtain similar results \cite{SFV16EPJ,DDB03PRC}. However, since the GPT form of the $nn$ interaction imparts less binding, a marginally higher strength of the three-body potential described in Eq. (\ref{3b}) is required to retrieve the same ground state energies \cite{SCH20PRC}.
	
	\begin{figure}[htbp]
		\centering
		\includegraphics[trim={0 0 0 0},clip,width=8.3cm]{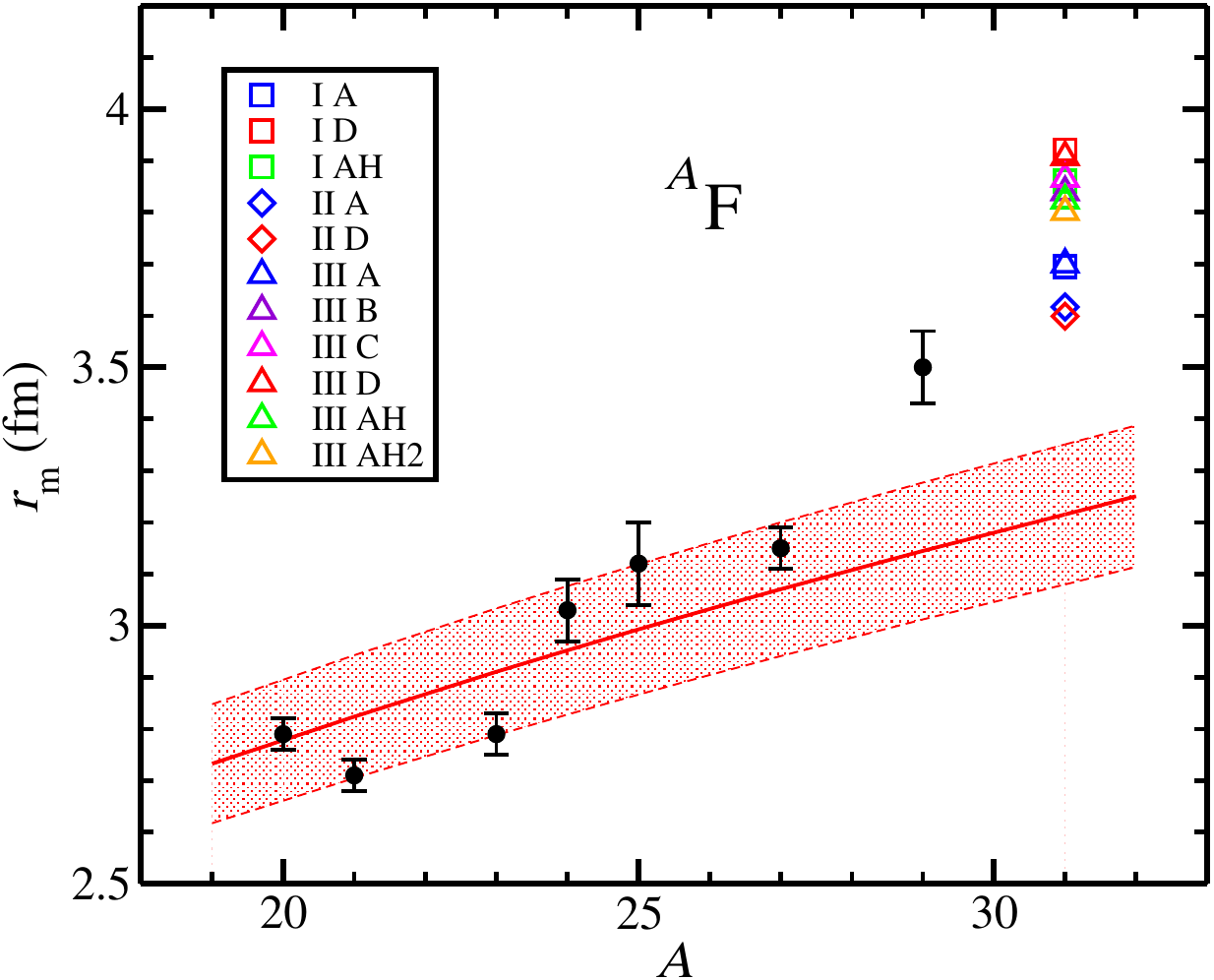}
		
		\caption{\label{fig: radii}  Variation of matter radius of Fluorine isotopes with mass number, \textit{A}. The experimental values, shown in black circles, for $^{20-25}$F isotopes were taken from Ref. \cite{OBC01NPA}, while those for $^{27,29}$F were extracted from Ref. \cite{BKT20PRL}. {The solid red line corresponds to the $R_0A^{1/3}$ fit for $A=20-29$, while the shaded area gives the region of 99.99\,\% confidence level in the fit.} Note the increase in the radii for $^{24}$F and $^{29}$F, as both are confirmed halos \cite{Abdullah20Pram, BKT20PRL, FCH20CP}. The squares represent case I, diamonds show case II and triangles display case III for the different values of $^{31}$F radius resulting from the different schemes under consideration. {The various colors portray the different sets, with blue depicting set A, violet for set B, magenta for set C, red for set D with the inverted schemes, green for anti-halo set AH and orange for AH2. Almost all configurations of $^{31}$F point to a two-neutron halo nucleus.}}
	\end{figure}
	In Fig. \ref{fig: radii} we show the variation of matter radius of Fluorine isotopes with increasing mass number. The experimental values for the radii for $^{20-25}$F were adapted from Ref. \cite{OBC01NPA}, while those for $^{27,29}$F were taken from Ref. \cite{BKT20PRL}. {The solid red line and the shaded area in the figure correspond to the weighted fit of the experimental data points for $A = 20-29$ and its prediction band up to 99.99\,\% confidence level. The fit is made with the standard $R_0A^{1/3}$ formula with an obtained $R_0=1.03$\,fm.} We do not show the radii for even isotopes having mass number, \textit{A} $>$ $25$ for the purpose of clarity.  It must be noted that the sharp rise of radius in mass number $24$ (compared to its bound predecessor) can be attributed to the fact that $^{24}$F is in itself, a one-neutron halo nucleus \cite{Abdullah20Pram}. 
	Also, the set of points representing the radii of $^{31}$F under the various working hypotheses in the present study does not include any ``theoretical error" and it will certainly be interesting to investigate and report the same in future analyses. 
	
	An examination of the numbers in Table \ref{T_obs} coupled with Fig. \ref{fig: radii} reveals vital new information. {It can easily be seen that radii of $^{29,31}$F lie much higher than the predicted $R_0A^{1/3}$ band in the figure.} 
	{Evidently,} this increase in the matter radius with increasing mass number of the Fluorine isotopes is consistent with the idea of $^{31}$F being a two-neutron halo, encompassing an established two-neutron halo core of $^{29}$F. No such case of a two-neutron halo within a two-neutron halo has been known so far in this island of inversion of the neutron rich medium mass region. Moreover, $^{29}$F is in itself, the heaviest known two-neutron halo, and it forming a core for a heavier as well as a drip-line nucleus to form another two-neutron halo opens a myriad of theoretical and experimental possibilities and challenges.
	
	Almost all the configurations considered in our study point to the sharp increase in the spatial extension, however, the exact numbers can be found in Table \ref{T_obs}. A comparison of $\Delta r$ manifests that some structural designs are more favorable for a halo {(with $\Delta r \gtrsim$ 0.30\,fm)} than others.  This value of $\Delta r$ has to be seen in light of the value for an established two neutron halo, $^{29}$F, where it was about 0.20\,fm \cite{CSF20PRC} (theoretical) and 0.35(8)\,fm \cite{BKT20PRL} (experimental) with respect to its $^{27}$F core. 
	
	In particular, all the sets with a considerable $p$-wave ground state contribution to the structure of $^{31}$F make it a halo nucleus, which is expected because it requires a low angular momentum centrifugal barrier to let the asymptotic of the valence nucleon(s') wave function to penetrate the classically forbidden region. This is also interesting in view of the anti-halo effect discussed by Masui \textit{et al.} \cite{MHK20PRC}, as there seems to be little proof of the large shrinking of the nucleus due to the large pairing energy when the $p_{3/2}$ and the $f_{7/2}$ lie closer than otherwise. In this observed effect, which can be attributed to the pairing energy, {when the energy gap is less than 0.4\,MeV, the valence neutrons should favour occupation to the $f_{7/2}$ level, resulting in a disappearance of the halo effect and hence, a smaller radius \cite{MHK20PRC}}. Indeed, from comparing sets D and AH in both cases I and III, the closer are the two energy levels, smaller is the matter radius. However, the {computed} decrease in the matter radius {in our study} is \textit{not substantial enough} so as to {prevent the formation of} a halo. Thus, the anti-halo effect is very small for this particular nucleus, but not insignificant. The largest halo possibility can be noticed in sets D of cases I and III, as those are the configurations that allow the maximum neutron occupation in the $p_{3/2}$ energy state while adding a neutron to $^{30}$F. This is also consistent with recent studies that promote a $p$-wave dominated g.s. two-neutron halo for $^{31}$F \cite{MLX20PRC,FR21arxiv}. A strong case of experimental verification, thus, emerges to firmly establish the two-neutron halo properties of $^{31}$F, {especially with regard to the anti-halo effect. In what follows, we would mostly concern ourselves with the inverted configurations due to their substantial $p$-wave contributions.} 
	
	Since the two-neutron separation energy of $^{31}$F has a significant uncertainty \cite{Wang17CPC} we also computed the matter radius for this nucleus at the extremes of the error limit. 
	These matter radius values for set D of case III are 4.084\,fm and 3.817\,fm corresponding to $S_{2n}$ values of 0.01\,MeV and 0.30\,MeV, respectively. It is evident that even at the higher end of the two-neutron separation energy, the matter radius is still large enough compared to its core $^{29}$F so as to result in a two-neutron halo nucleus at the Fluorine dripline.

	\begin{figure}[htbp]
		\centering
		\includegraphics[trim={0 0 0 0},clip,width=8.3cm]{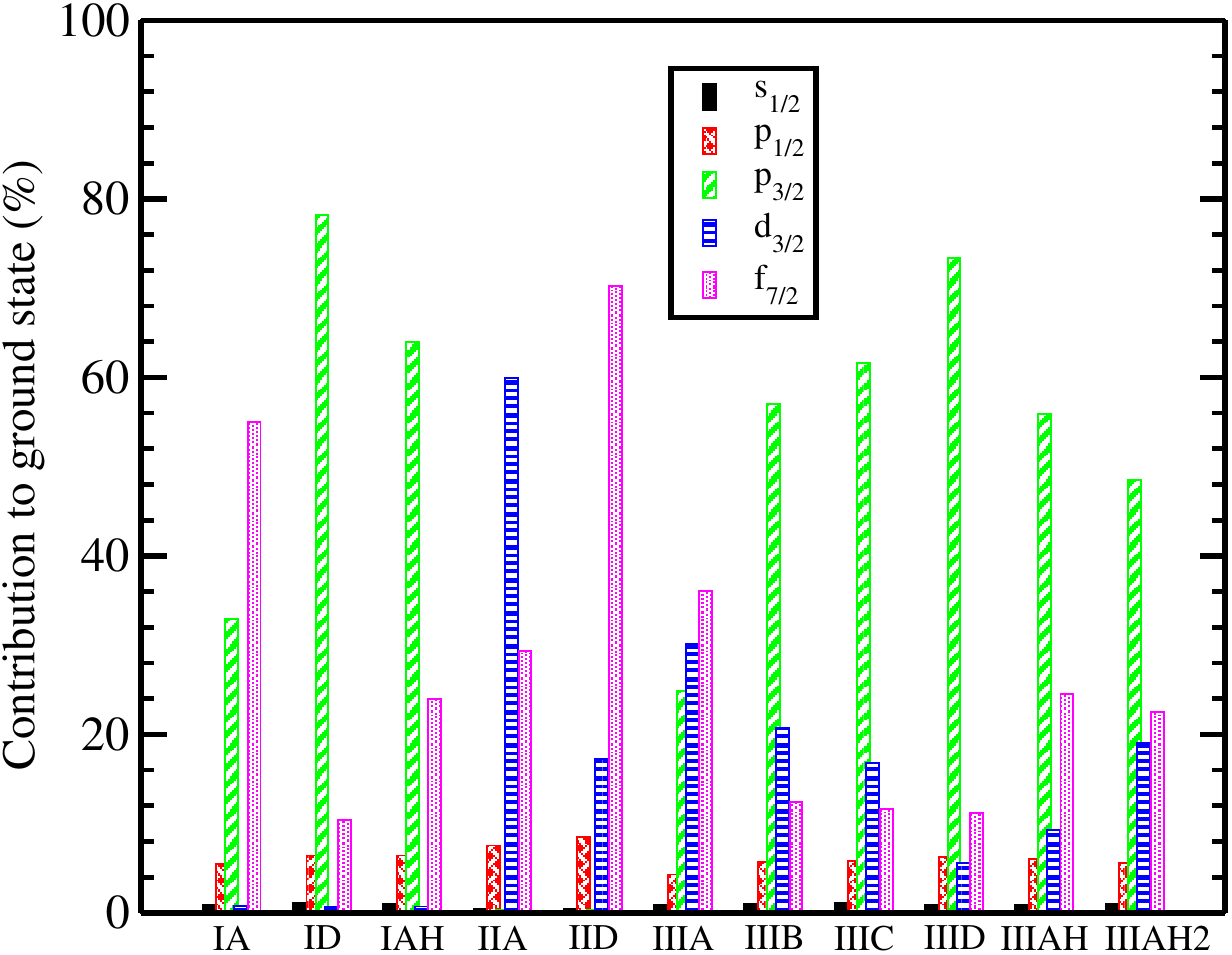}
		
		\caption{\label{fig: GScontri}  The contribution by different energy orbitals in the different configurations considered for the g.s. of $^{31}$F. The dominance of $p_{3/2}$, $d_{3/2}$ and $f_{7/2}$ subshells is evident for the various schemes.}
	\end{figure}
	
	Figure \ref{fig: GScontri} shows the percentage contribution to the ground state of $^{31}$F by the different energy orbitals in the various configurations considered in this study. This is extracted by transforming the system from the Jacobi-$T$ to the Jacobi-$Y$ set \cite{TND04CPC} and performing a change in the coupling order \cite{CSF20PRC}. The dominance of the different orbitals in this Manhattan plot is evidently a direct consequence of setting the resonance energies of the corresponding energy levels in the two-body potentials for the unbound $^{30}$F (cf. Table \ref{T_pot}) at their respective levels. Nevertheless, some interesting observations can be made. The contributions of the $s_{1/2}$, $d_{5/2}$ and $f_{5/2}$ levels is almost negligible (The histograms for the $d_{5/2}$ and $f_{5/2}$ levels were so small that they are not shown in the figure for the purpose of clarity). The same can also be said about the $d_{3/2}$ and $p_{3/2}$ contributions to cases I and II, respectively, which is expected since they are fully closed in their respective configurations and are Pauli forbidden. A subtle contribution of the $p_{1/2}$ level is visible throughout the histogram, which originates most probably from the non-resonant continuum of $^{30}$F. It can also be seen that all the sets that are halo candidates have very strong $p_{3/2}$ dominance in the ground state. Specifically, the inverted arrangements of (sets D) cases I and III have the highest $p_{3/2}$ influence, contributing about 78\% and 73\%, respectively. A concomitant analysis with Table \ref{T_obs} manifests that the higher the contribution of $p_{3/2}$, higher is the spread of the matter radius, consistent with the physics of halos being dominated by smaller orbital angular momentum configurations.
	
	Case III, set A, however, presents one of the most interesting cases.  Despite the resonance of the $p_{3/2}$ level lying much higher in energy than the $d_{3/2}$ and the $f_{7/2}$ levels, this open shell arrangement has a significant $p_{3/2}$ contribution to $^{31}$F ground state, \textit{pointing to the affinity of the nucleus} towards occupation of this low $\ell$ orbital in the open shell scenario. However, vital and comparable contributions from the $d_{3/2}$ and $f_{7/2}$ levels prevent it from supporting a substantially halo nucleus, something that can be corroborated from the matter radius variation in Fig. \ref{fig: radii}. In the anti-halo sets for cases I and III, the reduction in the energy gap between the $p_{3/2}$ and $f_{7/2}$ subshells ensures that the contribution of the $f_{7/2}$ increases substantially. However, we believe that the corresponding decrease in the matter radius is too small resulting in a small anti-halo effect, which is not sufficient to suppress the formation of a Borromean halo in $^{31}$F, as evident from the matter radius values and also the density distributions discussed below.

	\begin{figure*}
		\subfloat[]{\includegraphics[trim={2cm 0.5cm 1cm 0.9cm},clip,width = 2.4in]{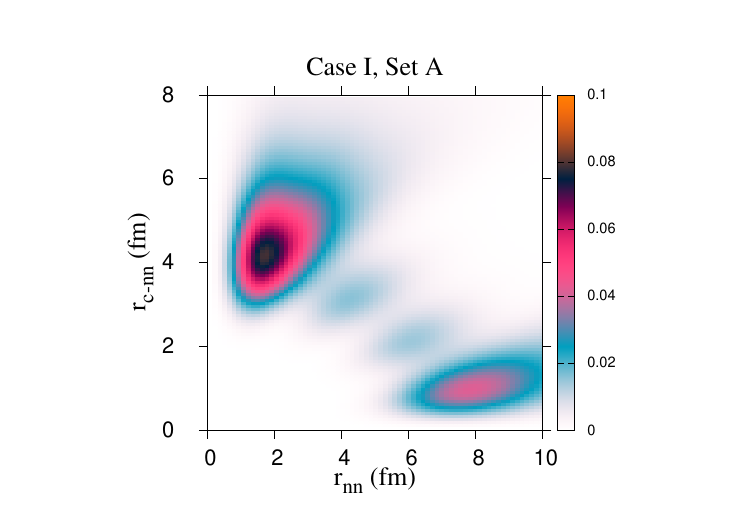}} 
		\subfloat[]{\includegraphics[trim={2cm 0.5cm 1cm 0.9cm},clip,width = 2.4in]{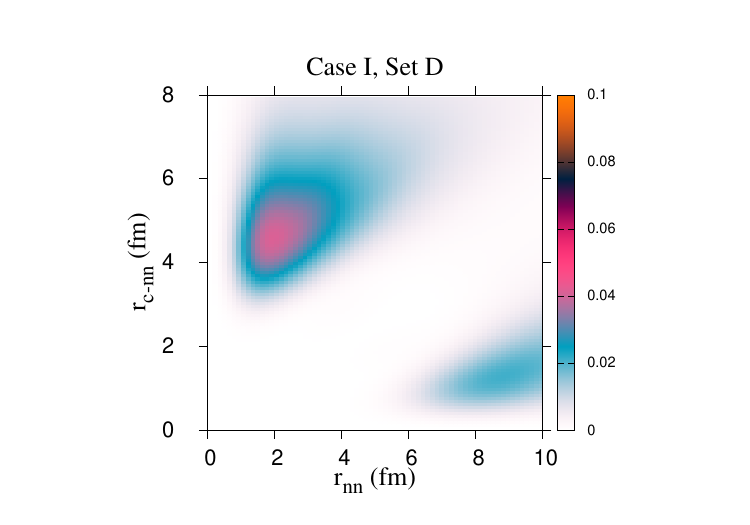}}
		\subfloat[]{\includegraphics[trim={2cm 0.5cm 1cm 0.9cm},clip,width = 2.4in]{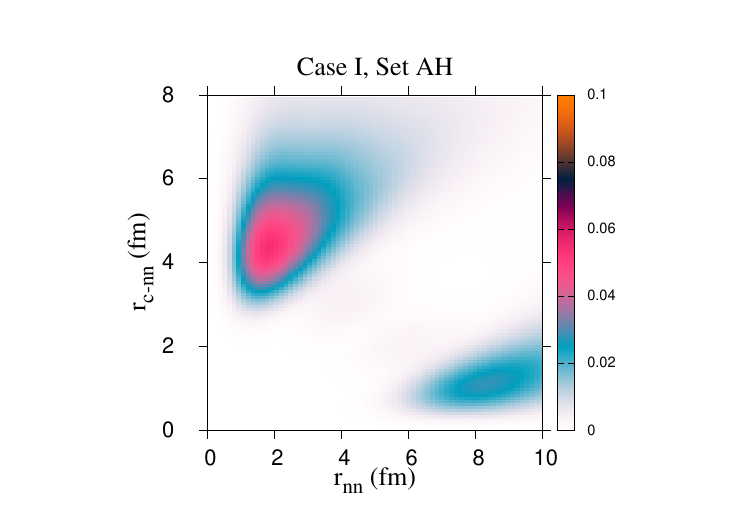}}\\
		\subfloat[]{\includegraphics[trim={2cm 0.5cm 1cm 0.9cm},clip,width = 2.4in]{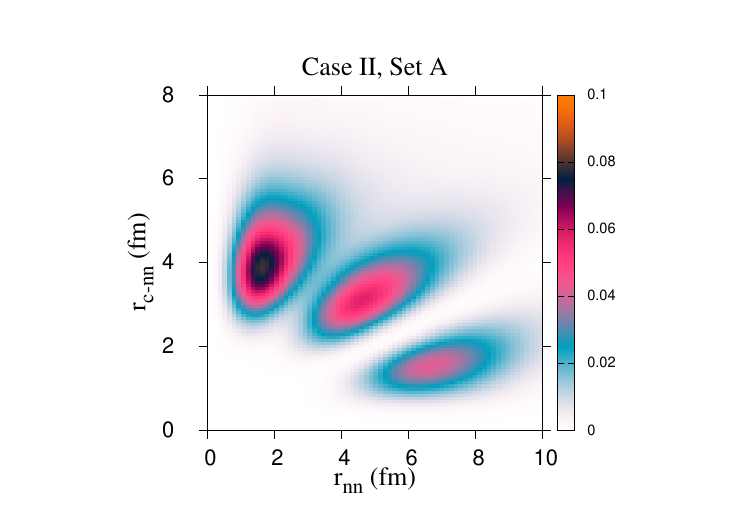}}
		\subfloat[]{\includegraphics[trim={2cm 0.5cm 1cm 0.9cm},clip,width = 2.4in]{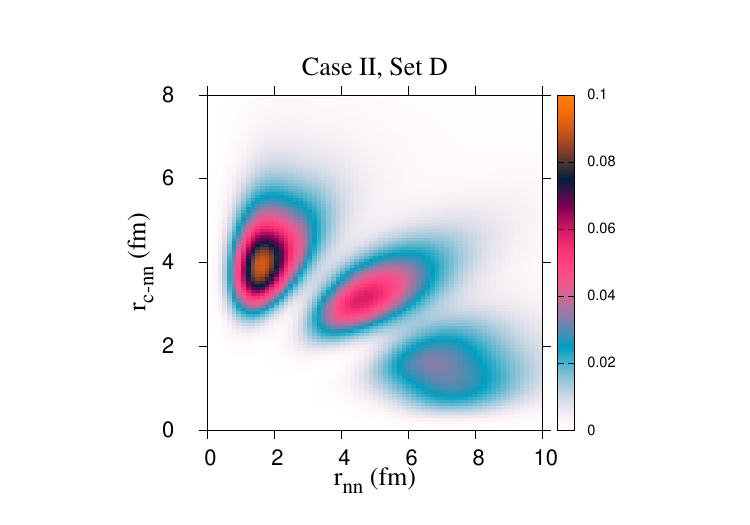}}\\
		\subfloat[]{\includegraphics[trim={2cm 0.5cm 1cm 0.9cm},clip,width = 2.4in]{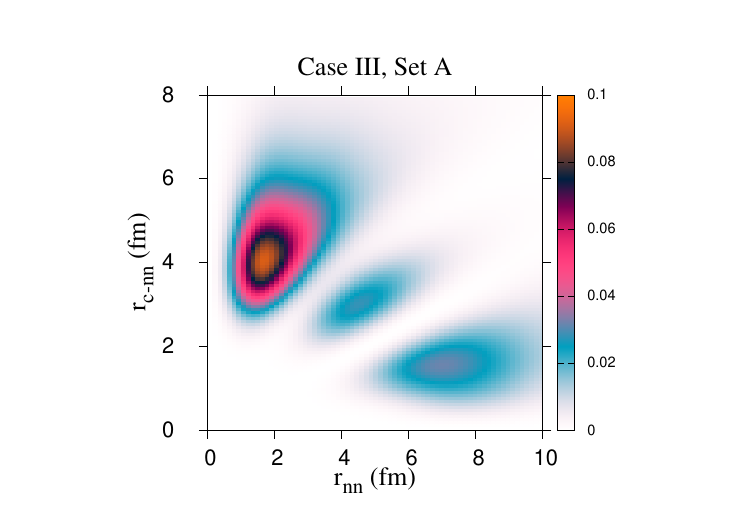}} 
		\subfloat[]{\includegraphics[trim={2cm 0.5cm 1cm 0.9cm},clip,width = 2.4in]{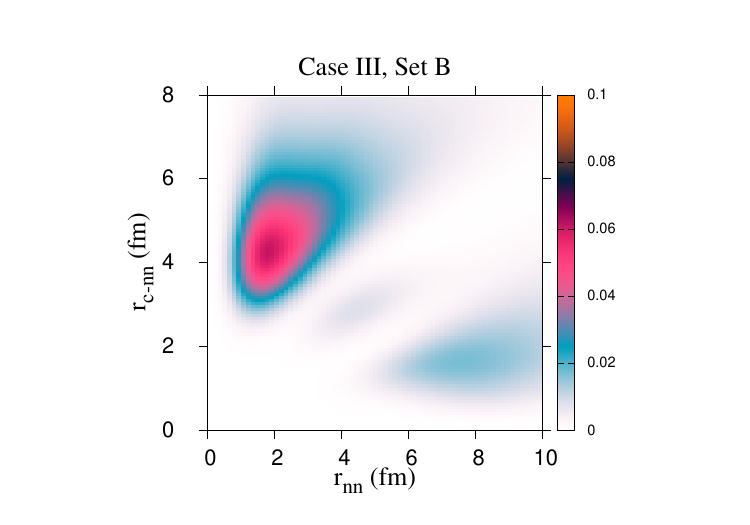}}
		\subfloat[]{\includegraphics[trim={2cm 0.5cm 1cm 0.9cm},clip,width = 2.4in]{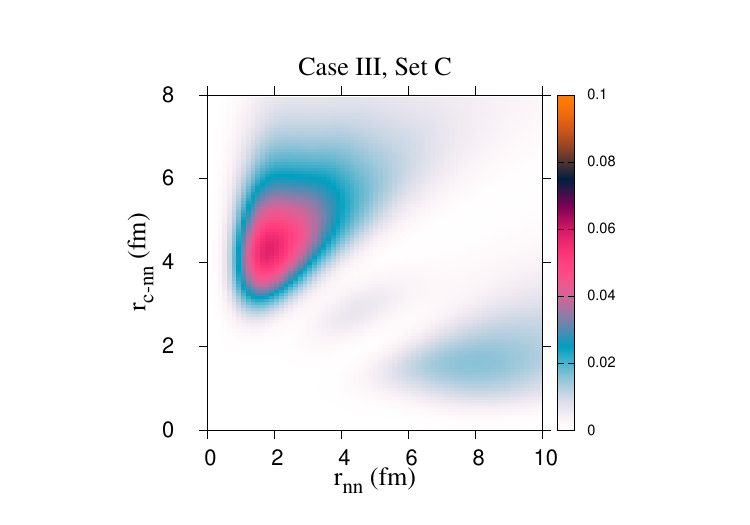}}\\
		\subfloat[]{\includegraphics[trim={2cm 0.5cm 1cm 0.9cm},clip,width = 2.4in]{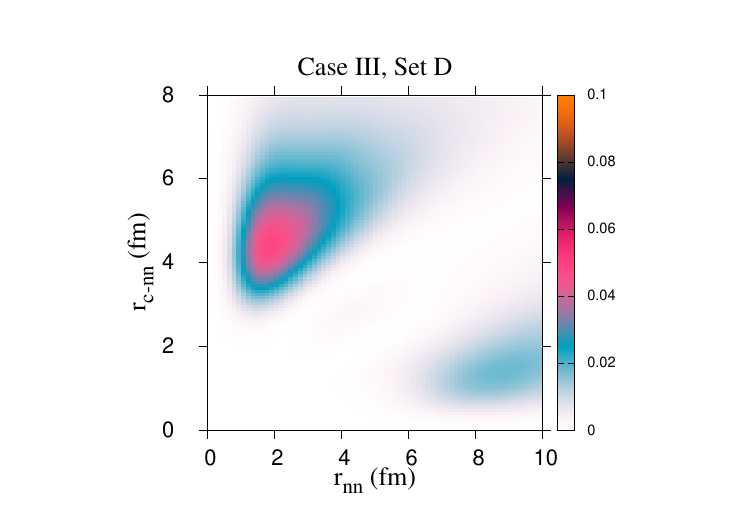}} 
		\subfloat[]{\includegraphics[trim={2cm 0.5cm 1cm 0.9cm},clip,width = 2.4in]{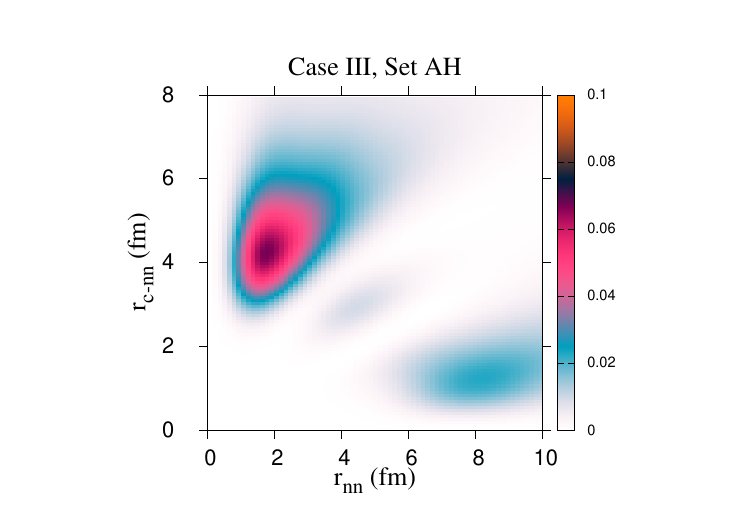}} 
		\subfloat[]{\includegraphics[trim={2cm 0.5cm 1cm 0.9cm},clip,width = 2.4in]{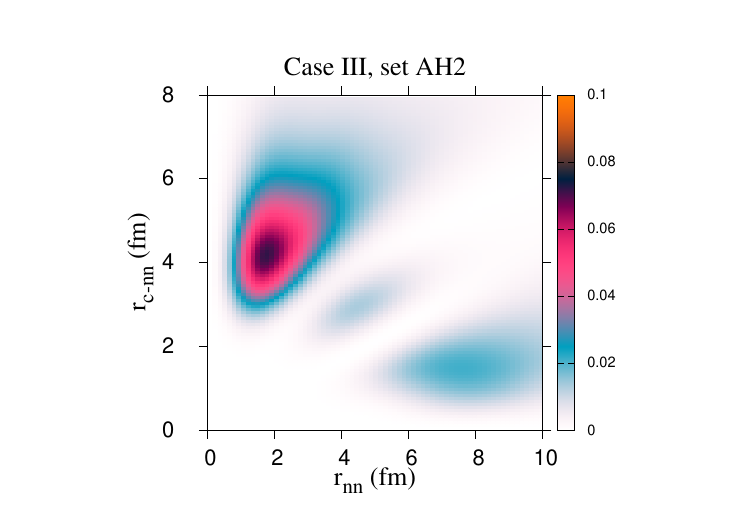}} 
		\caption{\label{fig: density} The ground state probability density distribution (in units of fm$^{-2}$) of $^{31}$F for all the configurations considered, as a function of $r_{nn}$ and $r_{c-nn}$.}
	\end{figure*}
	
	The density distributions for all the configurations are displayed in Figure \ref{fig: density}. The domination of the $d$-wave (especially in case II) is manifested through the presence of three peaks in the density spreads, with higher dominating configurations showing prominent amplitudes in the peaks. Case I is peculiar as there are four peaks observable in sets A and AH, pointing to the dominance of the $f$-wave. This $f$-wave contribution is also significant for other cases, however, the four-peak structure there is suppressed due to the stronger contributions by $d$- and $p$-waves, or in other words, lower angular momentum waves. The lowest of the three-peak structures arising from the $d$-wave is large enough to overshadow the two central peaks of the $f$-wave and thus, it is hard to disentangle them. 
	Further, the higher is the  $d$- or the $f$-wave component, higher is the amplitude of the peaks promoting the case of a mixed state for a smaller correlation angle between valence neutrons \cite{MLX20PRC}. 
	
	The \textit{dineutron} peaks \textendash or the peaks corresponding to the two correlated neutrons being spatially closer to each other for a given distance from the core (i.e., when $r_{nn}$ is small for any $r_{c-nn}$) \textendash in case III accumulate about three times the probability density than the opposite \textit{cigar-like} peaks (when $r_{nn}$ is large). In comparison, for cases I and II this ratio is always closer to one. This is a direct outcome of the larger mixing between the states of different parity due to the open shell configurations used in case III \cite{CIM84PRC, SCH20PRC, CNK20PRL, KCA20PRL}. However, the long density tail in each of the configurations in cases I and III for the dineutron peak is consistent for the case of other two-neutron halos \cite{SCH20PRC, HS05PRC, CKC19PLB}, being expectedly most smeared and spread for the inverted schemes. This extended tail causing the density blur is also evidence that
	the formation of a two-neutron halo in $^{31}$F at the given separation energy in the anti-halo configurations is still possible despite the increase in the pairing energy due to the nearing of the $f_{7/2}$ and the $p_{3/2}$ orbitals, {a result different than pointed out in Ref. \cite{MHK20PRC}}. The dineutron correlation is also significant when pointing to the low-lying dipole resonances of weakly bound nuclei \cite{NVS06PRL}, a feature we discuss in the next subsection. 
	\subsection{The dipole response}
	\label{sec: dipole}
	
	\begin{figure}[ht]
		\centering
		\includegraphics[trim={0 0 0 0},clip,width=8.3cm]{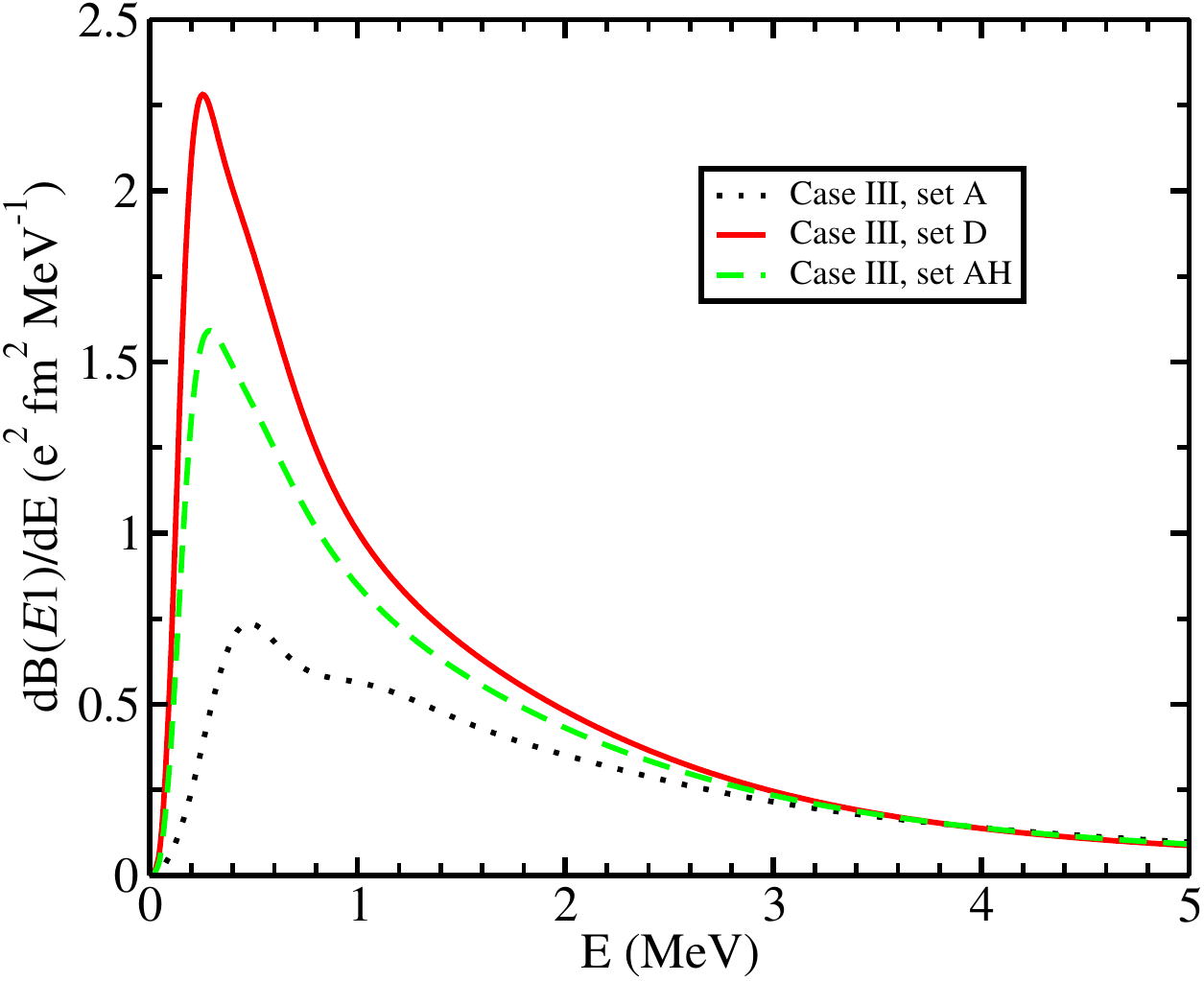}
		
		\caption{\label{fig: dBdE_caseIII}  The dB($E$1)/dE distribution for sets A (black dotted line), D (red solid line) and AH (green dashed line) of the open shell configuration (case III) considered in this study. Note the dominance of the inverted configuration, i.e., set D in the amplitude.}
		
	\end{figure}
	
	Electric dipole ($E1$) responses of exotic nuclei are good indicators of nuclear halos due to the peculiar soft dipole excitations and large B($E$1) strengths \cite{BS92PRC,NLV05EPJ,TB05NPA,MSS19EPJ}. {Theoretically, electric transitions can be described through a general expression of the reduced transition probability, }
	
	\begin{equation}
	\textrm{B}(E\lambda) = |\langle n_0j_0 ||\widehat{O}_{E\lambda}||n_1j_1\rangle|^2,
	\label{eq:BE1}
	\end{equation}
	{where $|n_0j_0\rangle$ is the ground state, $|n_1j_1\rangle$ represents each of the discrete excited states having a total angular momentum $j_1$ and $\widehat{O}_{E\lambda}$ is the electric operator of order $\lambda$. For a Jacobi-$T$ representation of a core + $n$ + $n$ system, this operator can be described via,}
	\begin{equation}
	\widehat{O}_{E\lambda} = eZ_cr_c^\lambda Y_{\lambda M_{\lambda}}(\widehat{y}).
	\label{eq:op}
	\end{equation}
	Here, $r_c$ is the relative distance of the core from the center of mass of the three-body system, given by $r_c=\sqrt{{2}/{A_c(A_c+2)}}y$ and $Z_c$ is the core charge.
	For a dipole transition, we set $\lambda$ = 1. 
	
	To analyse the electric dipole distributions of $^{31}$F, we select three cases of interest. We choose the open shell configurations as the $^{29}$F core itself is known to have open shell inverted scheme in its ground state \cite{CSF20PRC, BKT20PRL}. Keeping this case as our base, we set the inverted set D as the main candidate according to the contribution to the $^{31}$F ground state, while also comparing our results with set AH and selecting set A as a control.
	
	{The converged electric dipole responses for the three configurations of interest, up to a continuum energy of 5\,MeV, are presented in Fig. \ref{fig: dBdE_caseIII} for transitions between the g.s. $0^+$ and the continuum $1^-$ states. The curves are obtained by convoluting the discrete B$(E1)$ distributions with the Poisson distribution to obtain smooth, continuous distributions for the dipole transitions to the continuum \cite{RAG05PRC,CSF20PRC,CRA14PRC}. {More details about the dipole operator and the use of the Poisson distribution can be found in the Appendix.}} 
	
	{It is seen that maxima at energies of 0.5\,MeV, 0.28\,MeV and 0.32\,MeV are obtained for sets A (black dotted line), D (red solid curve) and AH (green dashed line) of Fig. \ref{fig: dBdE_caseIII}, respectively, indicating the vulnerability and dependence of the dipole strength on configuration mixing in the ground state of the nucleus as well as the radius. The transition curve for set D, when integrated cumulatively up to the depicted energy of 5\,MeV, gives an approximately similar value of 3.09\,$e^2$fm$^2$ for the dipole strength when compared to the sum rules (please see Appendix) listed in Table \ref{T_obs}. Other two curves of set A and AH follow a similar trend, in principle demonstrating that it is the low-lying response that constitutes a major strength of the distribution. The results approach the exact values when we integrate up to higher continuum energies of 10 or 15\,MeV. {A larger} amplitude {of} the dB($E$1)/dE curve {coupled with a} larger value of the B$(E1)$ distribution are another fingerprint of the enhanced spatial extension of the $^{31}$F nucleus in the inverted configuration prescribed through case III, set D. This can also be correlated through the values given in Table \ref{T_obs} that the higher the B($E$1) strength, larger is the radius promoting a halo structure. The results are, in fact, consistent with recent findings of higher B$(E1)$ values in other established two-neutron halos, $^{19}$B and $^{29}$F \cite{CSF20PRC, CNK20PRL}, increasing the probability of a two-neutron halo in $^{31}$F. The cumulatively integrated value for the anti-halo set, as can also be seen through a comparison from Table \ref{T_obs}, is smaller at 2.61\,$e^2$fm$^2$. Nevertheless, it is still large enough to indicate that the configuration mixing might not be too strong to demand and prevent the radial stretch of matter and might indeed result in a two-neutron halo.}
	
	{It is worth mentioning that similar features of low-lying dipole {strength} seem to be typical of two-neutrons halos like $^{19}$B, $^{11}$Li or $^{29}$F \cite{BS92PRC,CFR12PRL,CSF20PRC,NVS06PRL}. However, in the present calculations, a shoulder can be seen around 1\,MeV in the distribution for set A of case III in Fig. \ref{fig: dBdE_caseIII}. This could be attributed to the ambiguity in the smoothing procedure using the Poisson distribution \cite{CRA14PRC} and entails further investigations of the dipole continuum with {emphasis on possible resonances} and their widths, but goes beyond the scope of this work. Therefore, it is imperative that experimental measurements and analyses of the dipole distributions also be done for this dripline Fluorine nucleus so as to better constrain its structural properties.}	

	\section{Conclusions}
	\label{sec: Conc}
	
	In this work, we explore the possibility of the heavier known isotope of Fluorine, with mass number 31, for its possible structure and two-neutron halo feature. We treat it as a three-body structure and describe it using the hypershperical coordinates. We employ a pseudostate method to discretize its continuum using the analytical transformed harmonic oscillator basis for the purpose. In the calculations we assume $^{31}$F to be composed of an inert, spinless $^{29}$F core and two valence neutrons with a two neutron separation energy, $S_{2n}$, of 0.15\,MeV. We try and explore all the scenarios for the possible ground state configuration of $^{31}$F, including {mixed configurations (as encouraged by Ref. \cite{MLX20PRC}) and} the anti-halo arrangement {\cite{MHK20PRC}.} {We achieve this by fixing the core + $n$ potentials, and then adjusting the phenomenological three-body force to reproduce the exact two neutron separation energy. Computing different observables, like the matter radii, density distributions and dipole strengths, we} find that a strongly two-neutron halo structure is supported for all configurations with a significant and dominant $p$-wave contribution to its ground state, resulting in an increase in the matter radius $\gtrsim$ 0.30\,fm with respect to its core nucleus. This increase is impressive when seen in light of the corresponding increase for an established two-neutron halo like $^{29}$F \cite{CSF20PRC,BKT20PRL}. In particular, the arrangements where the $p_{3/2}$ level comes lower in energy than both the $d_{3/2}$ and $f_{7/2}$ levels favour the halo formation the most, with the inverted schemes enhancing the halo character beyond any doubt in the present analysis. The anti-halo configurations do result in a small reduction of the spatial extension owing to the increased pairing energy between the two valence neutrons {(due to the decreased energy gap between the $p_{3/2}$ and the $f_{7/2}$ energy states), however} the extent is not large enough to suppress the two-neutron halo formation. The analysis of the density distributions also points to the dominance of the $p_{3/2}$ contribution to the ground state of $^{31}$F in the open shell configurations.
	
	Calculations in the pseudostate continuum for the dipole distributions reveal a large $E1$ strength resulting from soft dipole excitations, another feature consistent with halo nuclei. Clear peaks at low energies above the core + $n$ + $n$ threshold in the computed d$B(E1)$/dE distributions and integrated strengths of more than 2.6\,$e^2$fm$^2$ for most of the configurations considered are a further evidence of a two-neutron halo formation in $^{31}$F as the numbers are similar to what one would expect while dealing with a proven case. However, a clear resolution and segregation of possibly resonant peaks was not possible for some of the arrangements like set A of case III and further investigations are required in that regard. In fact, this reaffirms the need for experimental measurements of the B$(E1)$ strength to constrain the halo structure of $^{31}$F.
	
	Moreover, additional data from transfer and knockout reactions involving this nucleus can shed light on the structure of this drip nucleus, that not only has a two-neutron halo with a strong $p$-wave contribution to its ground state, but also could be strongly deformed \cite{HAMA21PLB, LFL21PRC}. Our three-body wave functions would be handy in such a study as they include not only the bound, but the unbound states as well. Inclusion of possible core excitations as well as the core spin could also modify the results and would require a detailed level study of the $^{30}$F structure. 
	Thus, $^{31}$F presents an interesting case for future investigations at the edge of the nuclear chart that may provide insights into the interplay of dripline nuclear instability and {the} unusual phenomena appearing {in} the island of inversion.
	
	\begin{acknowledgments}
		This text presents results from research supported by SID funds 2019 (Università degli Studi di Padova, Italy) under project No.~CASA\_SID19\_01. 
		[GS] thanks Manuela Rodr\'{i}guez-Gallardo for discussing the numerical aspects used in this work. The authors also express their gratitude to Andrea Vitturi for insightful discussions.
	\end{acknowledgments}

	\appendix
	\section{Convergence of the results}
	\label{app1}
	\begin{figure}[htbp]
		\centering
		\includegraphics[trim={0 0 0 0},clip,width=8.3cm]{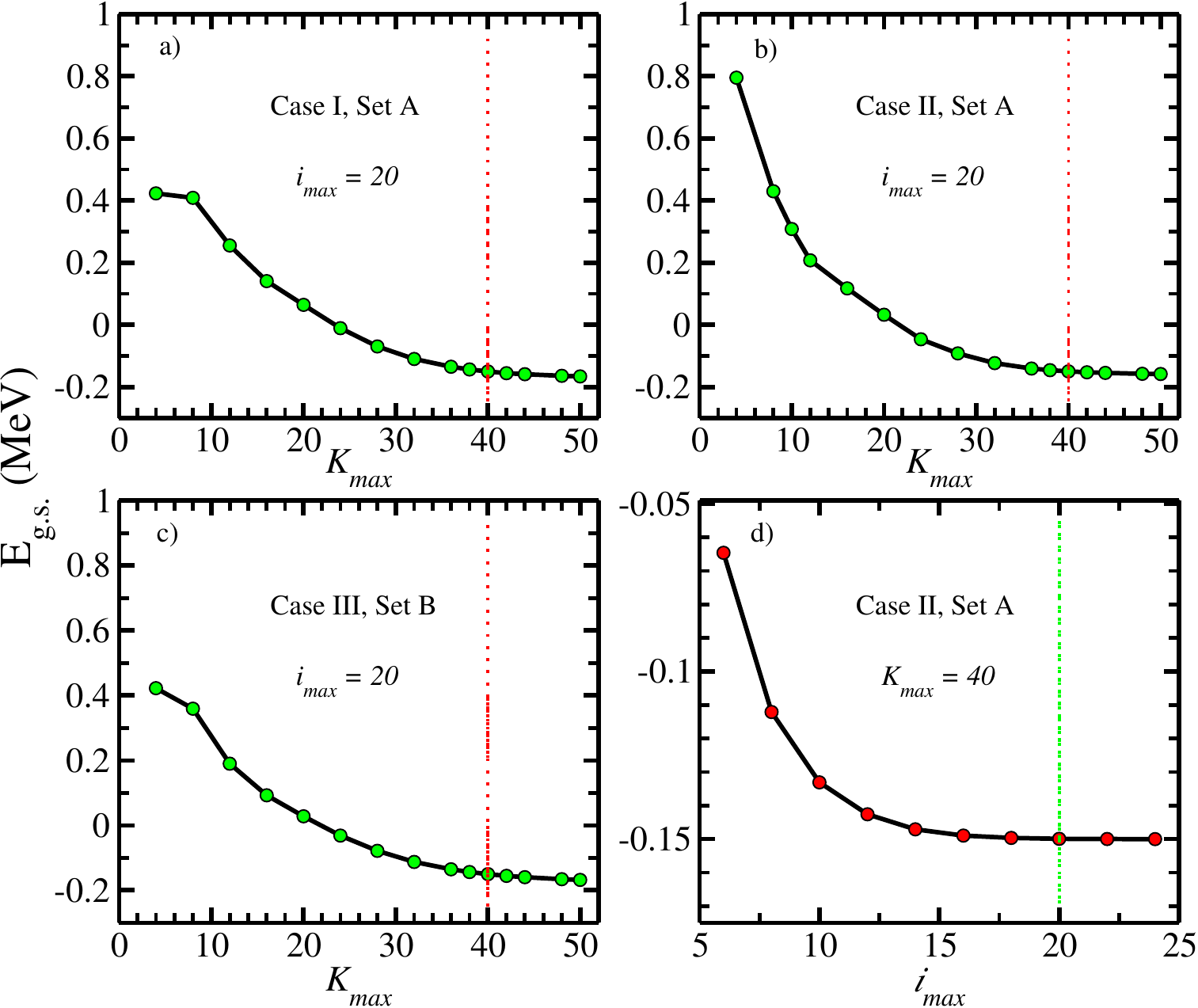}
		
		\caption{\label{fig: Con}  The convergence of the ground state two-neutron separation energy of $^{31}$F for $S_{2n}$ = 0.15\,MeV with variation in hyperspherical momentum $K_{max}$ and the basis size, $i_{max}$. Convergence is reached around $K_{max}$ = 40 and $i_{max}$ = 20. Calculations are done for Case I set A, case II set A and case III set B.}
	\end{figure}
	In generating the results presented in this manuscript, we had to ensure, of course, the convergence of our calculations. The position of the ground state energy converged with the size of the model space rather leisurely and evolved to a stable solution. Figure \ref{fig: Con} shows the convergence of the ground state energy as a function of the maximum hyperspherical momentum, $K_{max}$, and the basis size, $i_{max}$.  One of these was varied while the other was kept fixed to a sufficiently large value and although it is crucial to note that convergence depends on the choice of the basis, it required fairly larger values of these variables in comparison to previous works with the same basis \cite{CSF20PRC,CRA14PRC}, where also the calculations correspond to a THO basis with $b$ = 0.7\,fm and $\gamma$ = 1.4\,fm$^{1/2}$ for the ground state. For the present case, we achieved the desired level of accuracy at $K_{max}$ = 40 and $i_{max}$ = 20. Since the results are more prone to variations in the hypermomentum while the basis size reaches convergence fairly rapidly, we chose a set from each of the three cases to show the convergence tests with respect to $K_{max}$ for the g.s. energy, with set B in case III chosen specifically for the purpose of completeness. Fixing $K$ to a maximum value ensured that the orbital angular momenta corresponding to each Jacobi coordinate were limited to $l_x + l_y \leqslant K$ ({thanks to the properties} of the Jacobi polynomials in Eq. (\ref{eq:varphi})) and no additional truncation was required \cite{CSF20PRC}.
	
	\begin{figure}[htbp]
		\centering
		\includegraphics[trim={0 0 0 0},clip,width=8.3cm]{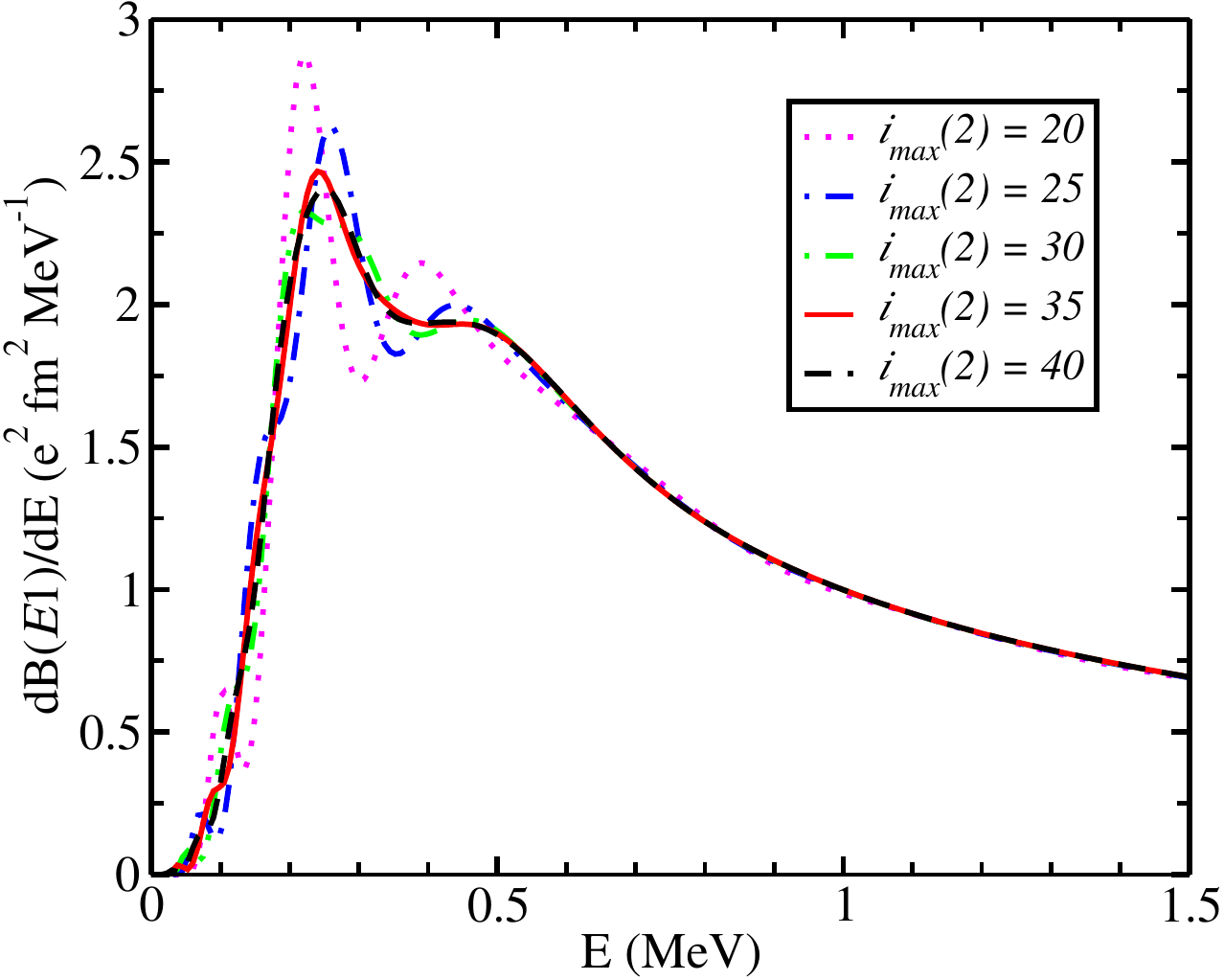}
		
		\caption{\label{fig: dBdE_i2max}  The dB($E$1)/dE distribution with variation in the basis size for the second state ($1^-$ in our case). The calculations converge at $i_{max}(2)$ = 35. The width of the Poisson distribution ($w$) used to generate these was fixed at 30, while the hypermomentum ($K$) used was 40. The calculations shown are done for case III, set D.}
	\end{figure}
	
	For the electric dipole distributions, we adjusted the size of the basis $i_{max}$(2) and also modified the value of $\gamma$(2) {(the 2 in the parentheses represents the continuum state for the dipole transition, with the ground state $0^+$ being the first) from 1.4 for the ground state to 1.0 for the excited states in the continuum, $1^-$. This was done because it enabled the convenient diagonalization of the continuum states Hamiltonian in a THO basis with a larger hyperradial extension.} The resultant reduction in the $\gamma/b$ ratio due to a decrease in $\gamma$ ensured a larger pseudostate density closer to the breakup threshold increasing the number of states, a desired trait for electromagnetic transition calculations \cite{CRA13PRC, Casal18PRC}. The convergence of the distributions was checked with the variation in basis size of the second state in steps of five, starting from $i_{max}$(2) = 20, and is shown in Fig. \ref{fig: dBdE_i2max}. The results are displayed for case III, set D but other configurations showed a similar behavior. It can be seen that convergent results are obtained for $i_{max}$(2) = 35. Hence, we used this value for the purpose of our calculations presented in this text. This completeness of the basis size is crucial to ensure that the calculated transition probabilities are comparable with the cluster sum rules for the electric dipole transitions \cite{CSF20PRC,DGJ08PRC}. The hypermomentum for the ground as well as the excited state, $K_{max}$, was kept fixed at 40 in all cases. {With these basis parameters, about 7300 energy states were obtained in the continuum between 0 - 5\,MeV.} Employing the same {interactions and the} three-body force also for the excited 1$^-$ states of the continuum as was used for the g.s. (0$^+$), we then searched for a stable B$(E1)$ distribution. The distributions were free from any further parameter fitting once this stable solution was reached.
	
	\section{Reduced transition probability and sum rules}
	\label{app2}
	
	{The electric dipole operator for a core + $n$ + $n$ system under the Jacobi-$T$ coordinate representation was extracted from Eq. (\ref{eq:op}) and for a $\lambda$ = 1 transition is,}
	
	\begin{equation}
	\widehat{O}_{E1} = eZ_cr_c Y_{1 M_{1}}(\widehat{y}).
	\label{eq:op1}
	\end{equation}
	
	{Then using Eqs. (\ref{eq:3bwf}) \& (\ref{eq:Upsilon0}), the electric dipole matrix elements can be written for three-body system comprising of a spinless core as \cite{CSF20PRC,CRA13PRC},}
	
	\begin{equation}
	\begin{split}
	\langle \text{g.s.}||\widehat{Q}_{E1}||n1\rangle = &\sqrt{3}Z_cer_c \sum_{\beta,\beta'} \delta_{l_xl_x'} \delta_{S_xS_x'} (-)^{S_x+l_x} \hat{l}\hat{l}'\hat{l}_y\hat{l}'_y \\
	& \times W(ll'l_yl_y';1l_x) W(01ll';1S_x) \begin{pmatrix} l_y & 1 & l_y' \\ 0 & 0 & 0 \end{pmatrix}\\
	& \times \sum_{ii'} C_\text{g.s.}^{i\beta 0} C_\text{n$_1$}^{i'\beta' 1}I_{i\beta,i'\beta'},
	\end{split}
	\label{eq:meQ}
	\end{equation}
	
	{where $\hat{\ell}=\sqrt{2\ell+1}$ and $W$'s are the Racah coefficients. The $I_{i\beta,i'\beta'}$ is the double integral of $y=\rho\cos\alpha$ between the hyperangular functions $\phi_{K}^{l_xl_y}(\alpha)$ of Eq. (\ref{eq:varphi}) and the hyperradial basis functions $U_{i\beta}(\rho)$ represented by Eq. (\ref{eq:THO}). Thus, $\widehat{Q}_{E1}$ is the electric operator for dipole transitions, which evidently is non-trivial for a three-body core + $n$ + $n$ case, but is analytic \cite{CRA13PRC,CSF20PRC}.}
	
	The non-energy-weighted sum rules on the other hand, for the study under consideration, depend only on the g.s. and can be defined as,
	
	\begin{align}
	S_T(E1) & = \sum_{n}\textrm{B}(E1)_{(\text{g.s.}\rightarrow n_1)} = \sum_n \left|\langle \text{g.s.}||\widehat{Q}_{E1}||n1\rangle\right|^2 \nonumber \\
	& = \frac{3}{4\pi}\frac{2Z_ce^2}{A_c(A_c+2)} \langle \text{g.s.}|y^2|\text{g.s.}\rangle,
	\label{eq:sumrule}
	\end{align}
	where the index $n_1$ labels the final states in a discrete representation \cite{CSF20PRC, CRA13PRC}.
	
	Within the present formalism, as is evident from the above equations, it is possible to extract the discrete B$(E1)_{(\text{g.s.}\rightarrow n_1)}$ values even though the $1^-$ pseudostates lie in the continuum. From these discrete B$(E1)$ quantities, one can fabricate an energy distribution through a standard convolution with a Poisson or a Gaussian distribution \cite{CSF20PRC,RAG05PRC,SFV16EPJ,Des20PRC} while preserving the total strength. The curves shown in Fig. \ref{fig: dBdE_i2max} are drawn for a distribution using the width of the Poisson distribution \cite{CRA14PRC}, $w$ = 30. For most of the configurations studied in this work including those shown in Fig. \ref{fig: dBdE_caseIII}, smoothing was achieved with a width of $w$ = 10. The smoothing procedure is required to remove the unphysical oscillations in the transition curves and must be adjusted to an adequate and optimum value so as not to broaden the dB($\pi\lambda$)/dE curves too much while at the same time, freeing the distributions from any unphysical peaks \cite{CRA14PRC}. Care must be taken with decreasing the width too much, as one might be tempted to smooth out the peaks which could also result from hidden and/or overlapping resonances. {Nonetheless, while the smoothing procedure may make the peaks broader or narrower, the total strength of the distributions remains unaffected.}
	
	{For a given PS having $\textrm{E}_n$ as the energy of the superposition of continuum states in the vicinity, a normalised form of the Poisson distribution can be defined for each discrete value of B$(E\lambda)(\textrm{E}_n)$ in the form \cite{CRA13PRC},}
	
	\begin{equation}
	\mathscr D(\textrm{E}, \textrm{E}_n, w) = \dfrac{(w+1)^{(w+1)}}{w!}\dfrac{\textrm{E}^w}{\textrm{E}_n^{w+1}}exp\left(\frac{-(w+1)\textrm{E}}{\textrm{E}_n}\right).
	\end{equation}
	
	{Then, one can easily convolute the discrete B($E\lambda$) distribution with $\mathscr D(\textrm{E}, \textrm{E}_n, w)$ to obtain the continuous dB($E\lambda$)/dE via \cite{CRA13PRC},}
	
	\begin{equation}
	\dfrac{\textrm{dB}(E\lambda)}{\textrm{dE}} (\textrm{E},w) = \sum_{n}\mathscr D(\textrm{E}, \textrm{E}_n, w)\textrm{B}(E\lambda)(\textrm{E}_n),
	\end{equation}
	{which is the distribution shown in Fig. \ref{fig: dBdE_caseIII}.}
	
	
	
	\bibliography{gaganbiblio}
	\bibliographystyle{apsrev4-1}
	
\end{document}